\documentclass[a4paper,UKenglish,cleveref, autoref, thm-restate, nolineno]{gd-lipics-v1}

\hideLIPIcs  

\title{The Perception of Stress in Graph Drawings} 

\author{Gavin J. Mooney}{Monash University, Australia \and \url{http://www.gavjmooney.com} }{gavin.mooney@monash.edu}{https://orcid.org/0009-0001-6208-4268}{}

\author{Helen C. Purchase}{Monash University, Australia \and \url{https://research.monash.edu/en/persons/helen-purchase}}{helen.purchase@monash.edu}{https://orcid.org/0000-0001-6994-4446}{}

\author{Michael Wybrow}{Monash University, Australia \and \url{http://users.monash.edu/~mwybrow/} }{michael.wybrow@monash.edu}{https://orcid.org/0000-0001-5536-7780}{}

\author{Stephen G. Kobourov}{Technical University of Munich, Campus Heilbronn, Germany \and \url{https://www.professoren.tum.de/en/kobourov-stephen} }{stephen.kobourov@tum.de}{https://orcid.org/0000-0002-0477-2724}{}

\author{Jacob Miller}{Technical University of Munich, Campus Heilbronn, Germany \and \url{https://jacoblmiller.github.io/homepage} }{jacob.miller@tum.de}{https://orcid.org/0000-0002-0567-785X}{}

\authorrunning{Gavin J. Mooney et al.} 

\Copyright{Gavin J. Mooney CC-BY} 



\category{} 

\relatedversion{} 

\acknowledgements{}

\EventEditors{John Q. Open and Joan R. Access}
\EventNoEds{2}
\EventLongTitle{42nd Conference on Very Important Topics (CVIT 2016)}
\EventShortTitle{CVIT 2016}
\EventAcronym{CVIT}
\EventYear{2016}
\EventDate{December 24--27, 2016}
\EventLocation{Little Whinging, United Kingdom}
\EventLogo{}
\SeriesVolume{42}
\ArticleNo{23}

\usepackage{algorithm}
\usepackage{algpseudocode}
\usepackage{url}
\usepackage{subcaption}
\usepackage{amsmath}

\usepackage{tabularray}
\usepackage{graphicx}
\usepackage{subcaption}
\usepackage{float}
\usepackage{pdfpages}

\usepackage{todonotes}

\usepackage{cite}
\usepackage{hyperref}
\hypersetup{
    colorlinks=true,
    linkcolor=blue,      
    urlcolor=blue,
    citecolor=blue,
}

\begin{document}

\maketitle

\begin{abstract}

Most of the common graph layout principles (a.k.a. ``aesthetics'') on which many graph drawing algorithms are based are easy to define and to perceive. For example, the number of pairs of edges that cross each other, how symmetric a drawing looks, the aspect ratio of the bounding box, or the angular resolution at the nodes. The extent to which a graph drawing conforms to these principles can be determined by looking at how it is drawn---that is, by looking at the marks on the page---without consideration for the underlying structure of the graph. A key layout principle is that of optimising `stress', the basis for many algorithms such as the popular Kamada \& Kawai algorithm and several force-directed algorithms. The stress of a graph drawing is, loosely speaking, the extent to which the geometric distance between each pair of nodes is proportional to the shortest path between them---over the whole graph drawing. The definition of stress therefore relies on the underlying structure of the graph (the `paths') in a way that other layout principles do not, making stress difficult to describe to novices unfamiliar with graph drawing principles, and, we believe, difficult to perceive. We conducted an experiment to see whether people (novices as well as experts) can see stress in graph drawings, and found that it is possible to train novices to `see' stress---even if their perception strategies are not based on the definitional concepts.

\end{abstract}


\section{Introduction}
\label{sec:introduction}

Algorithms for depicting graphs are based on conformance to one or more `layout principles' (or `aesthetics'). These principles are chosen so that the resulting drawing is considered `good', in terms of being easier to read or understand when compared to a random layout that does not take any useful principles into account. Some empirical work has been undertaken to determine whether these principles really do enhance human understanding, with support found for reducing the number of edge crossings \cite{goos_which_1997, purchase_graph_2012}, depicting symmetry \cite{goos_which_1997}, and having wide edge crossing angles \cite{huang_improving_2013}.

Many of these principles can easily be related to the process of visual perception. For example, the principle that edges adjacent at a node should subtend as wide an angle as possible relates to the limitations of visual acuity, as does the need to keep adequate space between nodes and edges; the principle of depicting symmetric subgraphs in symmetric form refers to the Gestalt law of symmetry; bent edges are contrary to the Gestalt law of continuity; an aspect ratio that is extremely different from the Golden Ratio can result in a sense of visual imbalance.

Optimising `stress' in a graph drawing (e.g. by~\cite{kamada_algorithm_1989,Gansner2005,zheng2019}) is a common layout principle that is much-cited by graph drawing researchers. It relates to the extent to which the geometric distance between each pair of nodes in the graph drawing is proportional to the shortest graph-theoretic path between them---over the whole graph drawing. The definition of stress therefore relies on calculations associated with the underlying structure of the graph (the `paths') in a way that other layout principles do not.

So, while most layout principles can clearly relate solely to visual perception, `stress' cannot---it requires analysis of the graph structure as well. This does not mean that there has been no empirical work that refers to stress, simply that it has not done so with explicit use of the word `stress'. Marner et al. conducted an experiment where participants were asked to untangle large graphs on a wall-display, with the support of a novel algorithmic technique for moving several nodes at once~\cite{Marner2014}. They found that the drawings created by users did not have lower stress or fewer edge crossings when compared with the graphs drawn using the Fruchterman \& Reingold algorithm~\cite{fruchterman_graph_1991}, and they question the importance of these two metrics for creating graphs suitable for human understanding. Their instructions to their participants make no explicit reference to the property of stress.

Chimani et al. claim that `people prefer less stress and fewer crossings'~\cite{chimani2014people}. Their two-alternative-forced-choice experiment asked participants to choose their drawing of preference, with the stimuli carefully chosen with variations in both stress and edge crossings. While it was shown that participants tended to prefer the drawings with lower stress (57\%) and fewer crossings (65\%), the former is an implicit variable (unseen and unexplained to the participants) and the latter is explicit---that is, it is obviously seen.

Stress-preference experimental papers like these make use of the computational stress measure in their data analysis, but do not educate the participants about the concept of stress---it is treated as a `hidden' feature, that cannot be `seen' by participants. This paper represents the first attempt to determine whether people can `see' stress in a graph drawing when the concept is explained to them. This is important for graph perception experiments that attempt to determine the most effective (or most preferred, or most efficient) layout algorithms for human understanding. Stress is a key feature of many common algorithms---if we cannot be sure that participants understand what it is, how can we assess the extent to which they value it in graph drawings?

Anecdotally, graph drawing researchers often claim that they can `see stress' - i.e., that it is as immediately perceptible as other layout principles. While this may be the case for those fluent in graph drawing principles forms (who have an internally and possibly intuitive understanding of the stress principle), it is unclear that `stress' can be explained sufficiently well to non-experts that they too can perceive stress in a graph drawing.

In this paper, we explore whether it is possible to `see' a feature of a graph drawing that is implicitly defined by both graph structure as well as visual form. We conduct a series of human experiments, asking participants to distinguish between pairs of graph drawings, identifying which has lower stress. While our results necessarily depend on the quality of the explanation of stress given to novices, we find that even experts find the task challenging.


\section{Background}
\label{sec:background}

\subsection{Applications of Stress in Graph Drawing}


Kamada and Kawai~\cite{kamada_algorithm_1989} first introduced the idea of using stress in graph drawing in the late 80s, with the motivation that a good drawing should accurately represent something about the graph structure. In this case, that drawn distances be proportional to graph-theoretic distances. The Kamada-Kawai layout algorithm has seen several improvements such as stress majorization~\cite{Gansner2005,Gansner2013} and stochastic gradient descent~\cite{zheng2019}. These most recent versions are among the most popular layout algorithms today, as can be seen by their implementation in libraries like NetworkX (kamada\_kawai\_layout), GraphViz (Neato), yEd (Organic), and others. 
Stress-based optimisation of multiple graph layout aesthetics is considered by Ahmed et al.~\cite{Ahmed2020} and Devkota et al.~\cite{Devkota2019}.

Kruiger et al.~\cite{kruiger2017} and  Zhu et al.~\cite{DRGraph21} generalize the classical dimensionality reduction algorithm t-SNE~\cite{van2008visualizing} and include stress as one of the evaluation criteria. Other papers that optimise stress-related functions and evaluate stress include those by Zhong et al.~\cite{zhong2023} and Xue et al.~\cite{9904492}. 

Regardless of whether an algorithm explicitly optimises stress, it is very commonly used as a metric to evaluate the quality of graph drawings and layout algorithms. A recent survey of graph layout algorithms~\cite{di2024evaluating} shows that it is the third most common metric employed in the GD community, behind running time and number of edge crossings. Stress is used in the evaluation of graph layout algorithms in papers by Hong et al.~\cite{Hong2019} and Marmer et al.~\cite{Marner2014}, as well in dynamic graph layout methods, as in Simonetto et al.~\cite{Simonetto2018} and Arleo et al.~\cite{arleo2022}. Brandes and Pich~\cite{brandes2009} evaluate several graph layout algorithms based on how well they optimise stress and Welch and Kobourov show that stress can be used as an alternative measure of symmetry in graph layouts~\cite{welch_measuring_2017}.

\subsection{Stress Definitions}
\label{sec:background_stress_defintions}

The definition of stress is much older than the Kamada-Kawai algorithm, with roots in statistical analysis. Stress as it is known in the graph drawing community began with Torgerson, who proposed a technique now known as metric Multi-Dimensional Scaling (MDS)~\cite{torgerson1952multidimensional}. Torgerson aims to provide a low dimensional representation of a set of objects on which distances are given. Importantly, little restriction is given on where these distances come from, e.g. they may be from a traditional metric space $\mathbb{R}^d$, responses to a Likert scale, or indeed, graph-theoretic distances. However the distances are obtained, they are collected into a matrix $D \in \mathbb{R}^{n \times n}$, with the cell $d_{i,j}$ containing the distance between object $i$ and object $j$. MDS aims to find a matrix $X \in \mathbb{R}^{n \times 2}$ where $X_i$ represents a low dimensional coordinate of object $i$ such that distances between rows in $X$ are exactly the distances in $D$. This is not possible in general, so the deviation of these distances are measured as a function: 
\begin{equation}
    \label{eq:full-stress}
    \sum_{i<j} \frac{(||X_i - X_j|| - d_{i,j})^2 }
    {d_{i,j}^2}
\end{equation}
Note that this function is differentiable with respect to $X$, so gradient-based optimisation schemes can be employed to find a local minimum of the equation. 

Sometime later, Kruskal defined what is now known as non-metric MDS~\cite{kruskal1964multidimensional}, with the motivation that preservation of exact distances is often very difficult and too restrictive. In the non-metric variant, it is instead important to maintain the \textit{rank} or ordering of distances i.e. from each object, the object that is first, second, and third closest in the original space should still be first, second, and third closest in the projection respectively. 
This non-metric stress is defined with a deceptively similar function: 
\begin{equation}
    \label{eq:non-metric-stress}
 \sqrt{\frac{\sum_{i<j}(||X_i - X_j|| - \hat{d}_{i,j})^2}{\sum_{i<j} {||X_i - X_j||}^2}}
\end{equation}
However, the matrix $\hat{D}$ is not the original distance matrix from the input. Instead, a Shepard diagram~\cite{shepard1962analysis} is formed by forming a set of coordinates for each pair of objects $i,j$: $c_{i,j} = (||X_i - X_j||, d_{i,j})$. A Shepard diagram created from an ideal drawing would have all points lying exactly on a straight line, since this would correspond to all input distances being exactly all output distances. To measure the deviation from this line, Kruskal performs a monotonic regression with $\binom{n}{2}$ points to best fit the diagram. The matrix $\hat{D}$ is defined such that $\hat{d}_{i,j}$ is the distance in the $x$-coordinate to the fitted line from $c_{i,j}$. 

In terms of information, the metric stress of equation~\ref{eq:full-stress} is more restrictive than the non-metric stress of equation~\ref{eq:non-metric-stress}. However, both are measuring how well distances are maintained in the output. Non-metric stress has the additional advantage that the resulting number will be in the range 0-1 and is hence more suitable as an evaluation metric. Scale can impact normalised stress values and several scale-invariant stress measures are detailed in recent papers by Ahmed et al.~\cite{ahmed2024size}, Smelser et al.~\cite{smelser2024normalized}, and Wang et al.~\cite{wang2023smartgd}.


\section{Methodology}
\label{sec:methodology}

\subsection{Experimental Methodology}

Our overriding Research Question is ‘Can people see stress in a graph drawing?’ As a simple yes/no question, this is tricky to address directly: it does not make sense to simply ask a participant to look at a graph drawing and to state whether it ‘is’ or ‘is not’ stressed. We therefore address this question by investigating whether people can see the difference in graph drawings with different stress values; thus, the research question becomes “Can people see differences in stress between two drawings of the same graph?”

Our methodology was mostly exploratory and incremental. We first addressed the research question by explaining the concept of stress to novices, giving them some training on the task (with feedback), and collecting performance data (accuracy, time, confidence) on 45 trials - over three different graph sizes. Each trial consisted of a pair of drawings of the same graph, with the participants being asked to indicate which has lower stress.

The results of this initial experiment were very encouraging, and so our follow-up study attempted to see how the extent of training affects performance; in this case, the participants received the same explanation of stress, but did not receive feedback on the training questions. We also asked some of our graph drawing expert collaborators to do the experiment as well; they did not have any training session.

Section~\ref{sec:procedure} explains the experimental procedure in more detail.

\subsection{Measuring Stress}
\label{sec:stress_measure}

To test the perception of stress, we need a set of graph drawings which have different values of stress. To create these, we must define a measurement of stress which is independent of graph size, structure, and drawing space. While the general definition (i.e. the metric definition of equation~\ref{eq:full-stress} and its commonly used variants), can be used to calculate the stress of a given drawing, it does not meet this criteria. For example, using this definition, Welch and Kobourov show that the stress of a given drawing varies with respect to the scale of the drawing space~\cite{welch_measuring_2017}. This means graph drawings with different geometric scales cannot be compared.

To create the stimuli, we use the definition of stress provided by Kruskal~\cite{kruskal1964multidimensional} (equation~\ref{eq:non-metric-stress}). While intended for measuring the quality of multi-dimensional scaling techniques, the method is applicable to measuring stress in graph drawings and has numerous benefits. Firstly, the measure is normalised, and hence applicable to graphs of different size and structure. This definition also has the benefit of being independent of the rotation and geometric scale of the drawing space. To integrate this measure with other graph drawing metrics, such as those described by Mooney et al.~\cite{mooney2024multi}, we subtract the Kruskal stress value from 1. This means that, as a metric, a value of 1 represents the extreme that is intuitively assumed to be good---in this case, zero stress. For the rest of this section, we refer to this definition as the `Kruskal Stress Metric' (KSM). 

Hopefully, Section~\ref{sec:background_stress_defintions} leaves one convinced that Kruskal's definition is an appropriate measure of stress. To further justify this choice, we used the large dataset described by Mooney et al.~\cite{mooney2024multi}, and calculated the Kruskal stress of nearly half a million graph drawings, then compared this to a normalised stress function based on the Kamada and Kawai definition~\cite{kamada_algorithm_1989}. We found the correlation between the two definitions to be 0.871. This strong correlation highlights the similarity in the respective definitions.

\subsection{The stimuli}
\label{sec:stimuli}

We ran three concurrent experiments, for graphs with node count of 10, 25 and 50. For each graph size, we created five random graphs using the Erdős–Rényi model~\cite{erdos_renyi}, with the constraint that the graph is connected and the number of edges is less than two times the number of nodes. This constraint ensures that drawings do not end up as `hairballs'. Preliminary testing when creating the stimuli showed that denser graphs (and hence denser drawings) tend to have higher stress than sparser graphs, and so, inclusion of them would limit the range of possible stress values.

We create graph drawings with KSM values in the range of 0.4 to 0.8, with intervals of 0.05 (nine drawings in a set). For each graph size, we generate three sets of drawings using a basic hill climbing algorithm. The evaluation function for this algorithm is the absolute difference between the KSM of the current drawing and the target KSM value. Nodes are initially positioned randomly within a unit grid. Each iteration updates the drawing by selecting a random node, and moving this node to a new random position within the initial bounding box of the drawing space. This new random position is also bounded by a circle of decreasing radius (as the number of iterations increases) centred on the chosen node's current position. If the new drawing has KSM closer to the target value, it is taken to be the current drawing, otherwise it is discarded. The algorithm terminates when the KSM of the drawing is within 0.01 units of the target value. Example stimuli are shown in Table~\ref{tab:stimuli}.

\begin{table}[t]
\centering
\caption{Example drawings of different KSM values. Each row consists of different drawings of the same graph. Higher KSM values indicate lower stress.}
\label{tab:stimuli}
\begin{tblr}{
  cell{1}{3} = {c=3}{c},
  cell{2}{3} = {c},
  cell{2}{4} = {c},
  cell{2}{5} = {c},
  cell{3}{1} = {r=3}{},
  cell{3}{2} = {c},
  cell{4}{2} = {c},
  cell{5}{2} = {c},
  vlines,
  hline{1-3,6} = {-}{},
  hline{4-5} = {2-5}{},
}
              &    & \textbf{KSM} &     &     \\
              &    & 0.4          & 0.6 & 0.8 \\
\textbf{Size} & 10 & \includegraphics[width=0.25\textwidth]{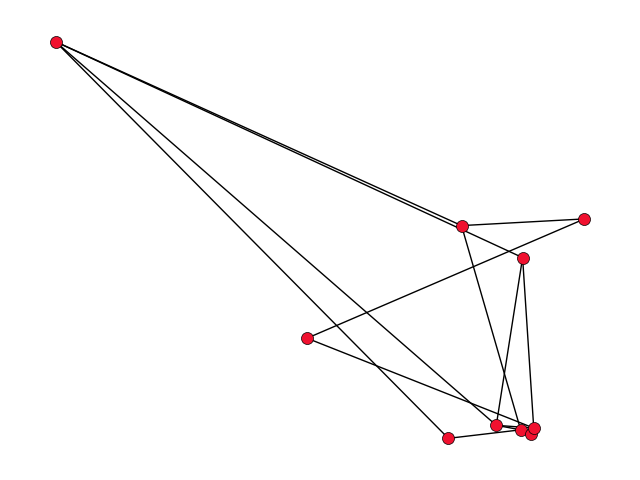} & \includegraphics[width=0.25\textwidth]{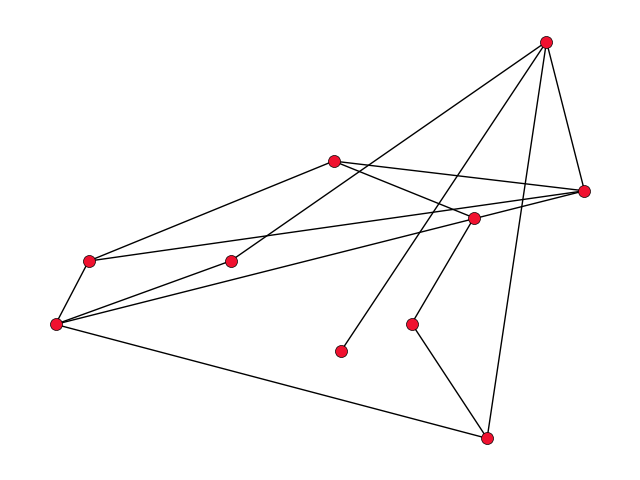} & \includegraphics[width=0.25\textwidth]{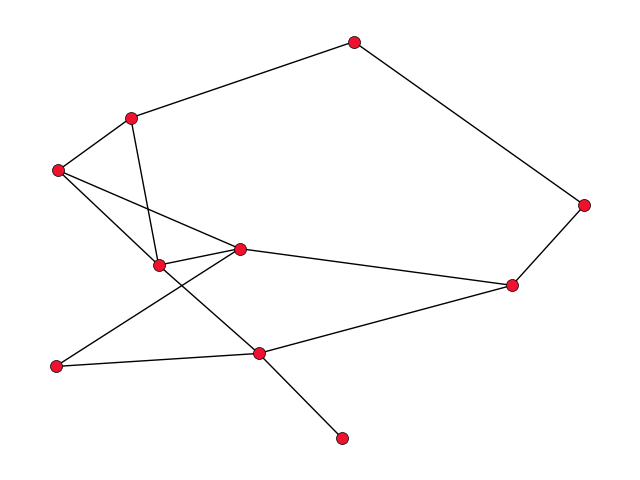} \\
              & 25 & \includegraphics[width=0.25\textwidth]{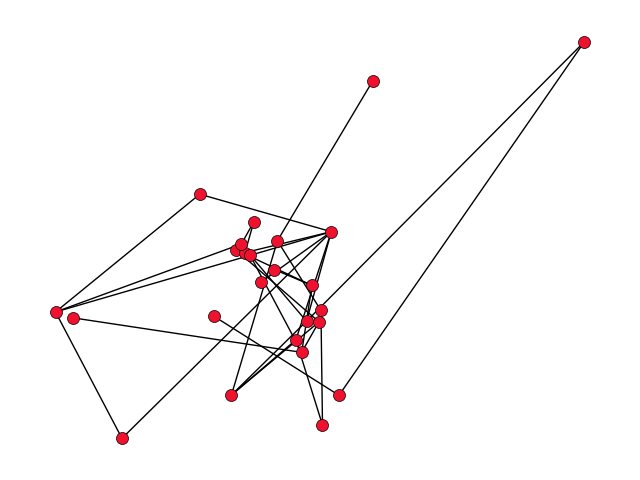} & \includegraphics[width=0.25\textwidth]{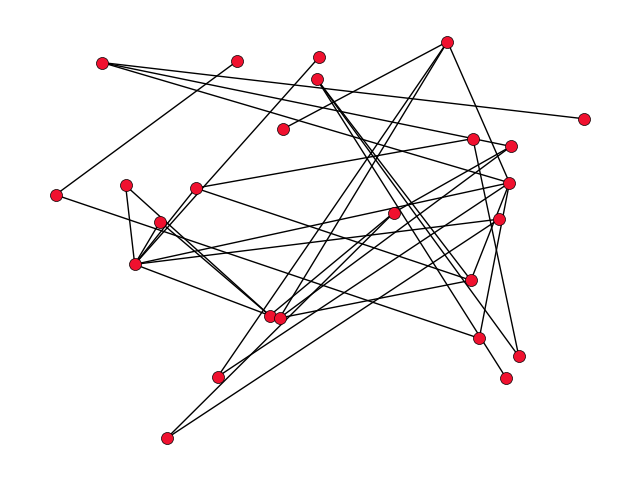} & \includegraphics[width=0.25\textwidth]{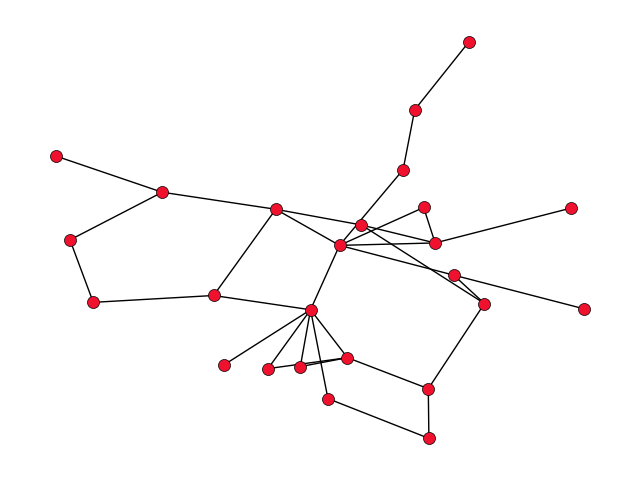} \\
              & 50 & \includegraphics[width=0.25\textwidth]{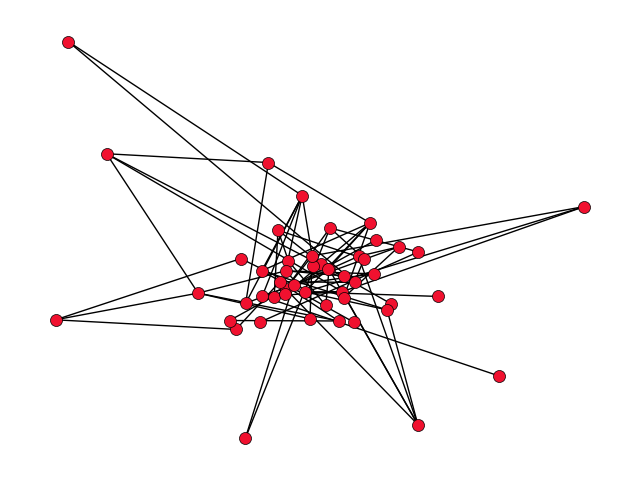} & \includegraphics[width=0.25\textwidth]{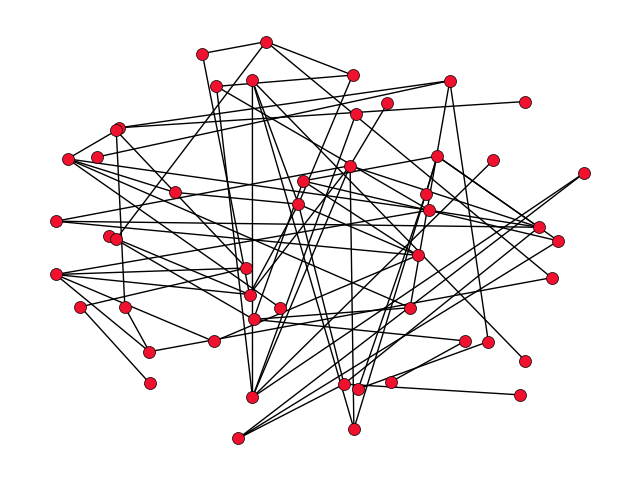} & \includegraphics[width=0.25\textwidth]{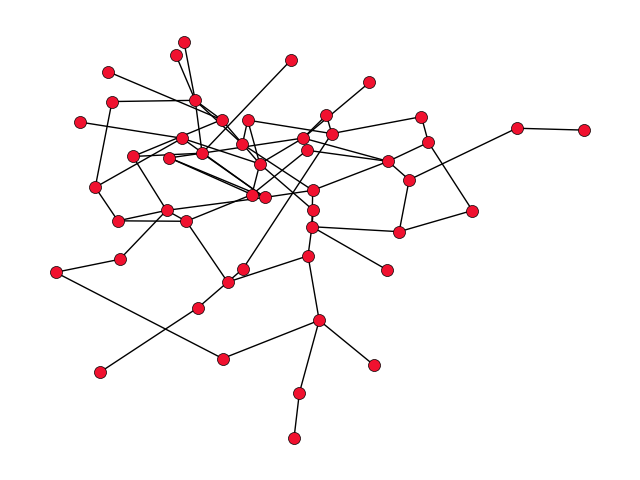}     
\end{tblr}
\end{table}

\subsection{Experimental Procedure}
\label{sec:procedure}
The methodology for each novice experiment is the same, but with different stimuli (corresponding to the three different graph sizes). The participants are first shown an overview of the aims of the experiment and asked for their consent to participate. Next, participants are shown a page which outlines the key concepts of graph drawings and stress (Appendix~\ref{sec:appendix_stress}). We use layman terms such as `network' instead of `graph', `object' instead of `node', and `connection' instead of `edge'. This page explains the following :
\begin{itemize}
    \item What a graph is, including definitions for nodes and edges.
    \item What a graph drawing is, and that one graph can have many different drawings.
    \item That different graph drawings have different visual properties, such as symmetry.
    \item What is meant by a `path' and `shortest path' between two nodes.
    \item What stress means in relation to graph drawing, with very basic examples of low and high stress drawings.
    \item Four pairs of drawings with low and high stress.
    \item That the participants should not spend a long time trying to calculate the stress of a drawing, but should rely on their immediate perception.
\end{itemize}

Participants are then shown nine predetermined pairs of drawings and asked to choose which has lower stress (or that they have the same stress). The nine drawings are of graphs that are the same size as the given experiment with differences of KSM ranging from 0.4 to 0, decrementing by 0.05 (in that order). In the first set of experiments there are 25 participants for each graph size (75 total). These 75 participants are given feedback immediately after each pair to inform them if they were correct or, if not, what the correct response was. Where feedback is given, over 50\% of the responses to these training questions must be correct to continue to the main experiment. In the second set of experiments there are 10 participants for each graph size (30 total). These 30 participants are given no feedback on the nine training questions. 

\begin{figure}[H]
    \centering
    \includegraphics[width=0.99\textwidth]{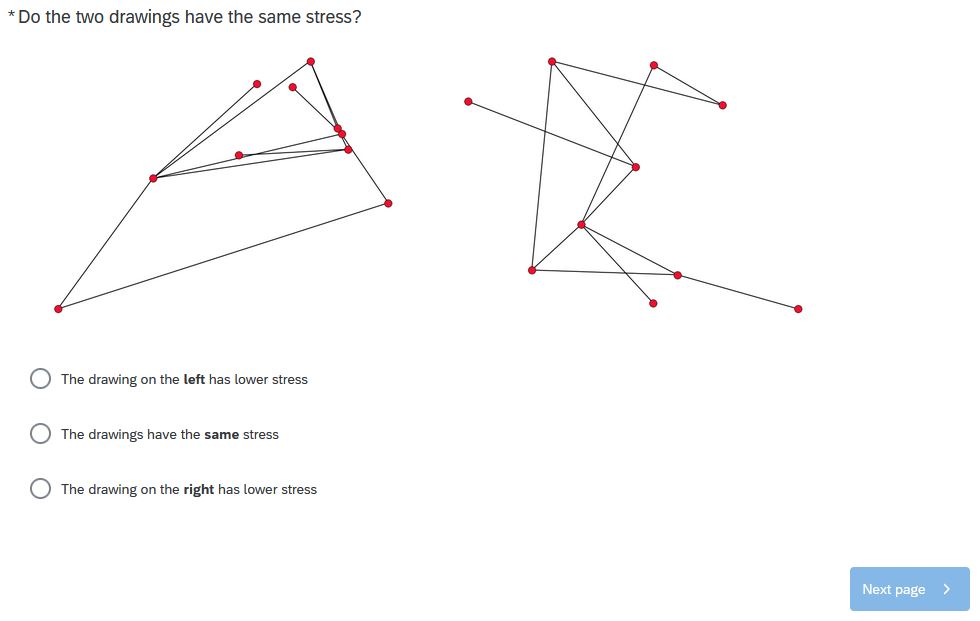}
    \caption{Example trial for the n=10 experiment. Here the drawing on the left has a KSM value of 0.45, while the right drawing has a KSM value of 0.7 (but this information is, of course, not shown to the participants).}
    \label{fig:trial}
\end{figure}

In the main part of the experiment, participants are shown 45 pairs of drawings (trials), with KSM values in the range 0.4--0.8, and asked to choose the drawing with lower stress (or that they have the same stress). Each pair consists of two drawings of the same graph. There are five graphs and nine unique KSM differences (0, 0.05, 0.1...0.4). The exact drawings for each KSM difference is random---e.g., for graph one, and KSM difference of 0.3, one participant may be shown drawings with KSM of 0.4 and 0.7, whilst another may be shown drawings with KSM of 0.5 and 0.8. The order in which pairs of drawings are shown is randomised for each participant, as well as which two drawings are chosen (from a pool of three) and the order (left/right) that they appear on the page. This randomness is incorporated to lower the chance of a learning effect. The participants are not given any feedback about their choice. After each choice, the participant is asked if they were confident about their choice. An example trial is shown in Figure~\ref{fig:trial}.

After all 45 pairs have been shown, the participant is asked some follow up questions about the experiment and asked for some demographic information. See Section~\ref{sec:data_collection} for more details.

The expert experiment follows the same methodology as the novice experiments, however the experts are not shown the nine training pairs and complete the 45 trials for each graph size (135 total). The experts are shown the 135 trials in blocks of 45, corresponding to each graph size (10, then 25, then 50), with an opportunity for a break between blocks.

\subsubsection{Data Collection}
\label{sec:data_collection}

In each experiment we collect the following data:
\begin{itemize}
    \item The number of incorrect training responses (out of nine).
    \item Which pairs of drawings were shown to participants and their KSM values, as well as the order they were shown in. Within each pair we also keep track of which drawing was displayed on the left and which was displayed on the right.
    \item The response from the participant about which drawing had lower stress (The drawing on the left has lower stress/The drawings have the same stress/The drawing on the right has lower stress).
    \item How confident the user was about each choice (Confident/Not confident).
    \item The time taken to submit an answer for each pair.
\end{itemize}

After all pairs of drawings are shown we also collect data for some additional questions and demographic information:

\begin{itemize}
    \item The overall strategy used to determine which drawing had lower stress.
    \item The participant's overall confidence in their responses (Very confident/Somewhat confident/Not very confident/Not confident at all).
    \item How difficult the participant found the experiment (Very difficult/Difficult/Easy/Very Easy).
    \item How familiar the participant is with network diagrams (Very familiar/Somewhat familiar/Not very familiar/They are new to me).
    \item The participants' age and gender.
\end{itemize}

\subsubsection{Experimental Conduct}
The online survey platform `Qualtrics' was used to set up and run the experiments, and collect the data. 

Participants were recruited from the online platform `Prolific' and paid at a rate of £9 GBP per hour. The pool of participants was limited to users over the age of 18 and residing in either the United Kingdom or Australia. Participants who were included in one study were excluded from participating in any of the others to reduce the chance of a learning effect.

For each graph size, we collected results from 25 participants who received feedback on the training, and 10 participants who did not. For the participants who received feedback on the training, those who answered more than 50\% incorrectly were excluded from the main experiment. For the experiments on graphs of size 10 and 25, only one participant did not pass the training. For the experiment on graphs of size 50, three participants did not pass the training. For each experiment, we recruited more participants until we had 25 who passed the training and completed the whole experiment. Each experiment was first piloted by two participants to ensure the experiments ran smoothly and that the data was being collected correctly. No changes were made after the successful pilots.

The expert participants were invited via email, of whom ten participated in the experiment. Two of these ten responses were incomplete, leaving us with eight complete expert responses.

This research was approved by the Monash University Human Research Ethics Committee with ID 42695.



\section{Results}
\label{sec:results}

\subsection{Trained Novices (TN)}
\label{sec:results1}

We address our research question (``Can people see the differences in stress between two drawings of the same graph?'') by analysing our data along two dimensions, giving two sub-questions:

\begin{itemize}
    \item RQa: ``How much does the quantifiable difference in stress (the `delta') between two drawings of the same graph affect the perception of stress?''
    \item RQb: ``How much does the size of the graph affect the perception of stress?''
\end{itemize}

In all cases, our dependent variables for measuring perception are accuracy, response time, and confidence. 
Of the 75 participants, there were 38 women, 36 men and 1 gender-diverse. The median age range was 36-45, with 16 participants between 18-25, and one over 76. Four participants said they were `very familiar' with graph drawing; 55 claimed novice status (at most `not very familiar').

\subsubsection{Delta trends}
\label{sec:results2}

\begin{figure}[H]
    \centering

    \includegraphics[width=0.49\textwidth]{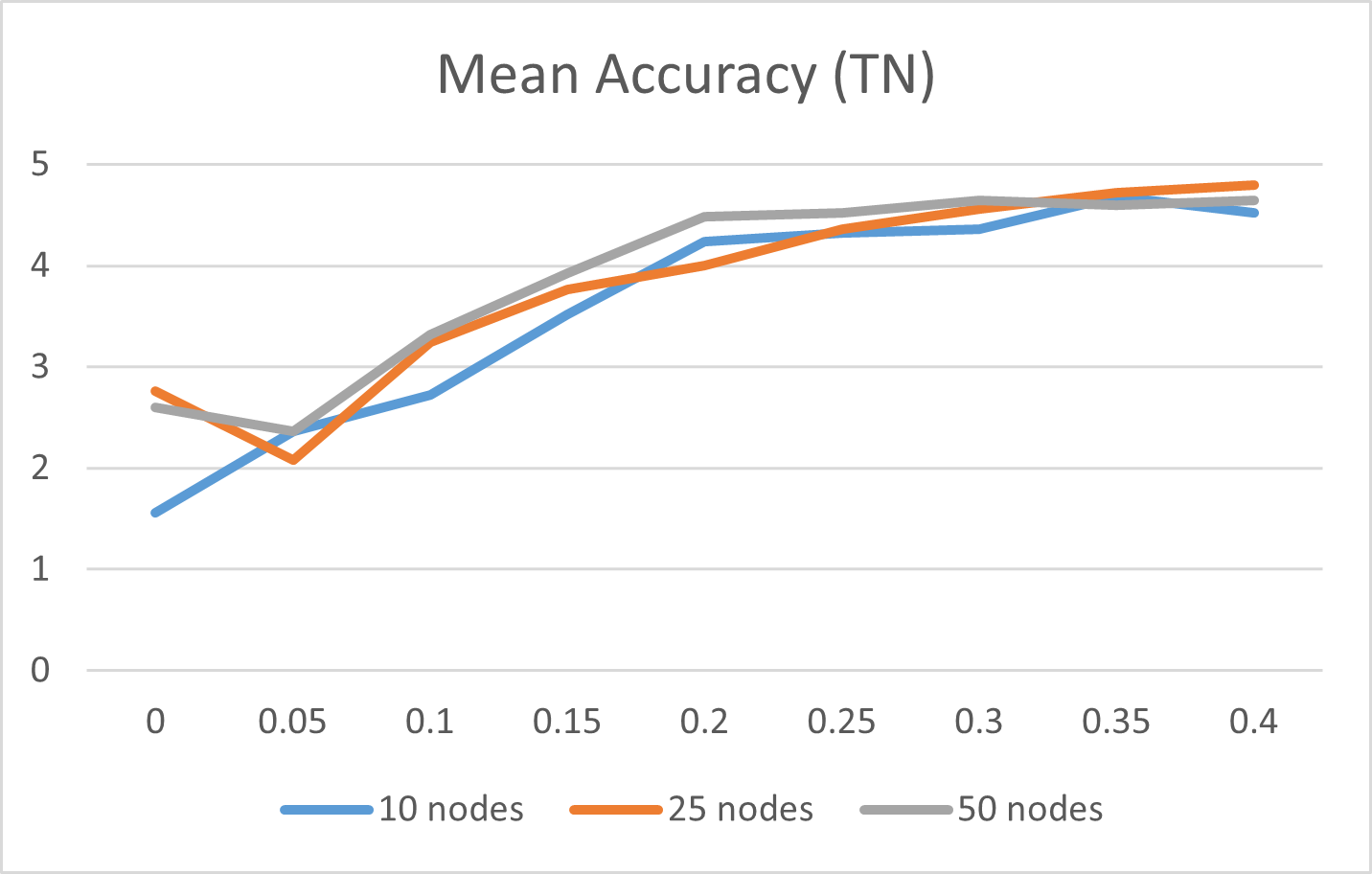}
    \includegraphics[width=0.49\textwidth]{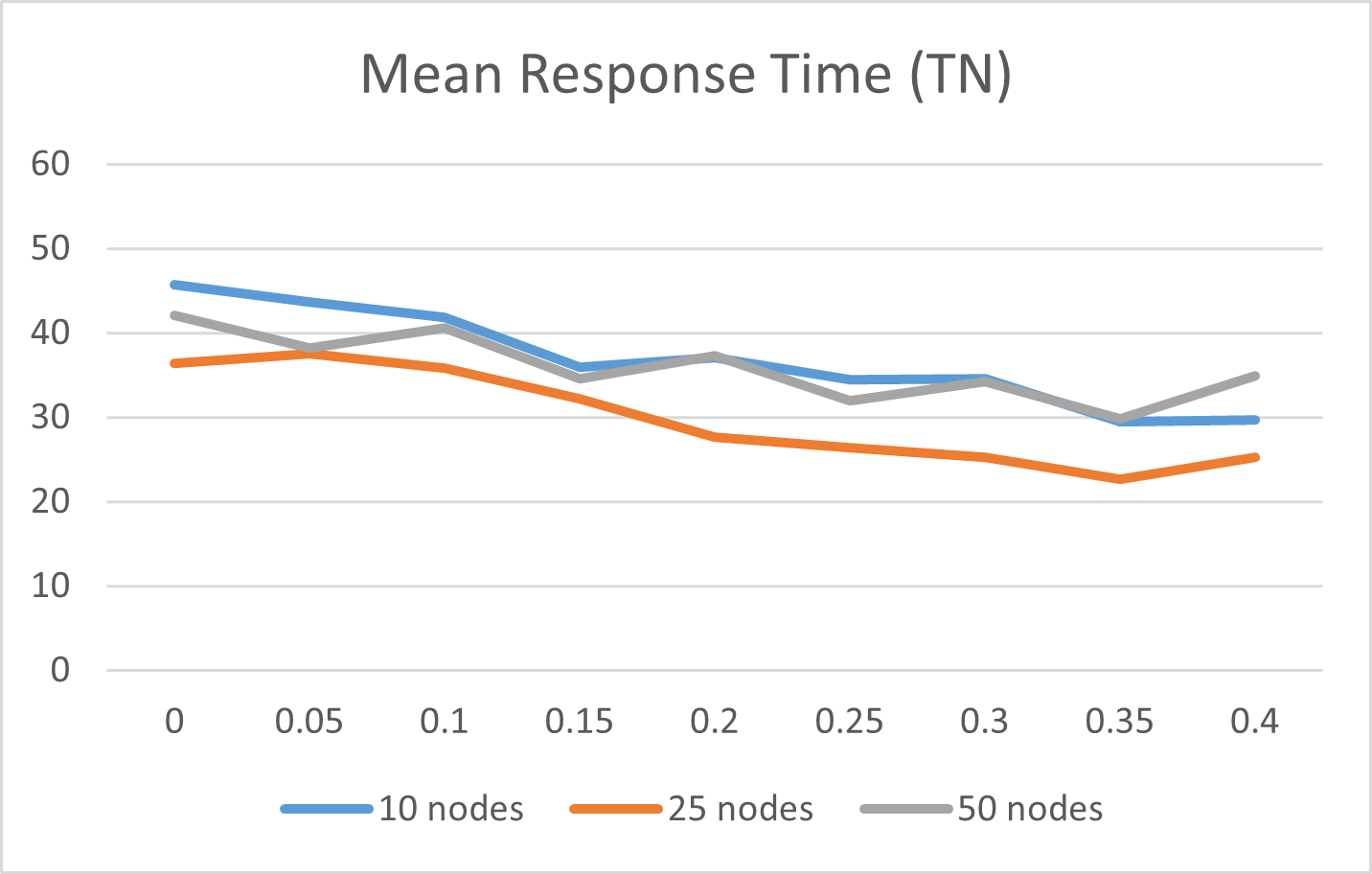}

    \includegraphics[width=0.49\textwidth]{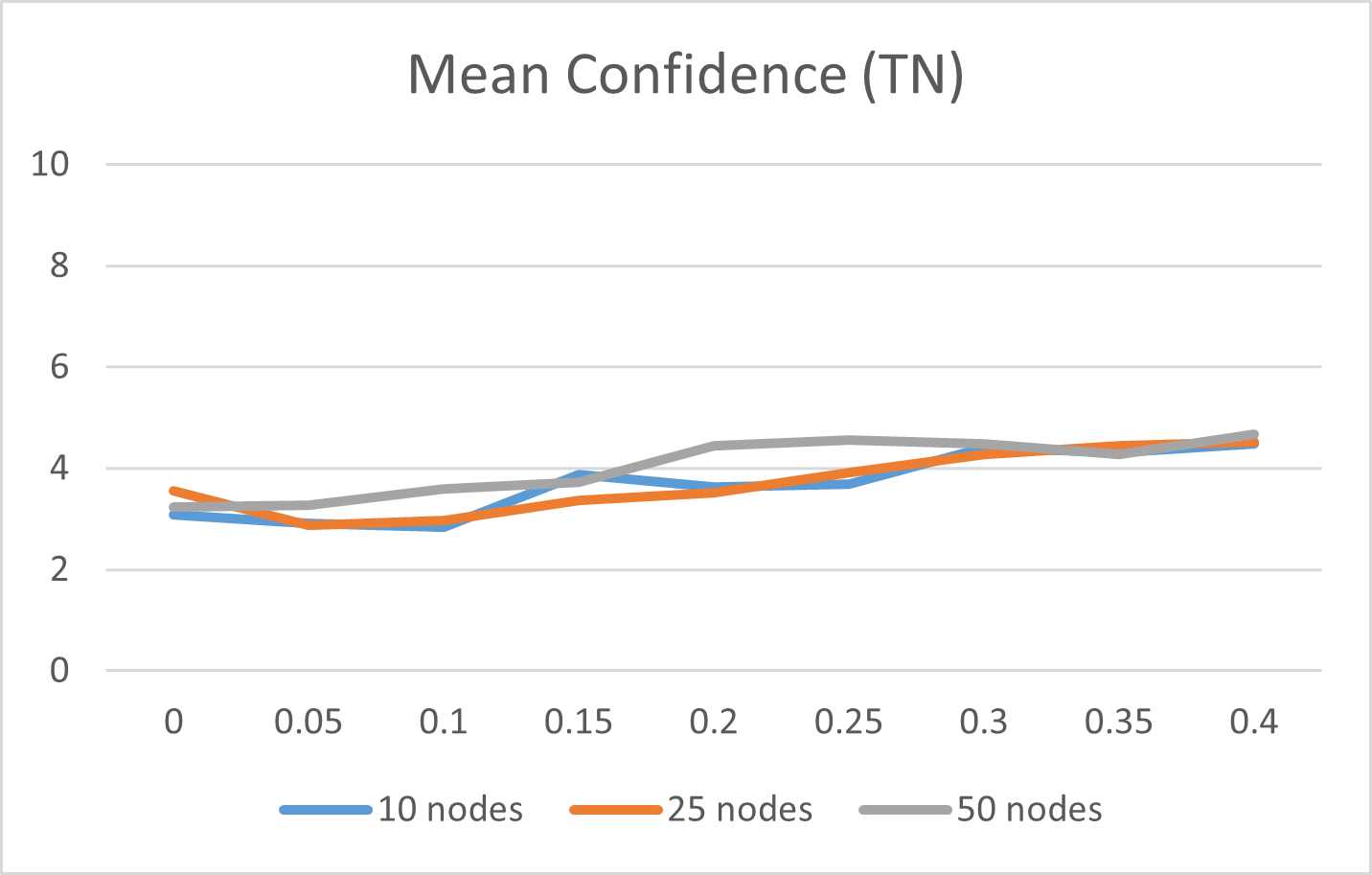}
    \includegraphics[width=0.49\textwidth]{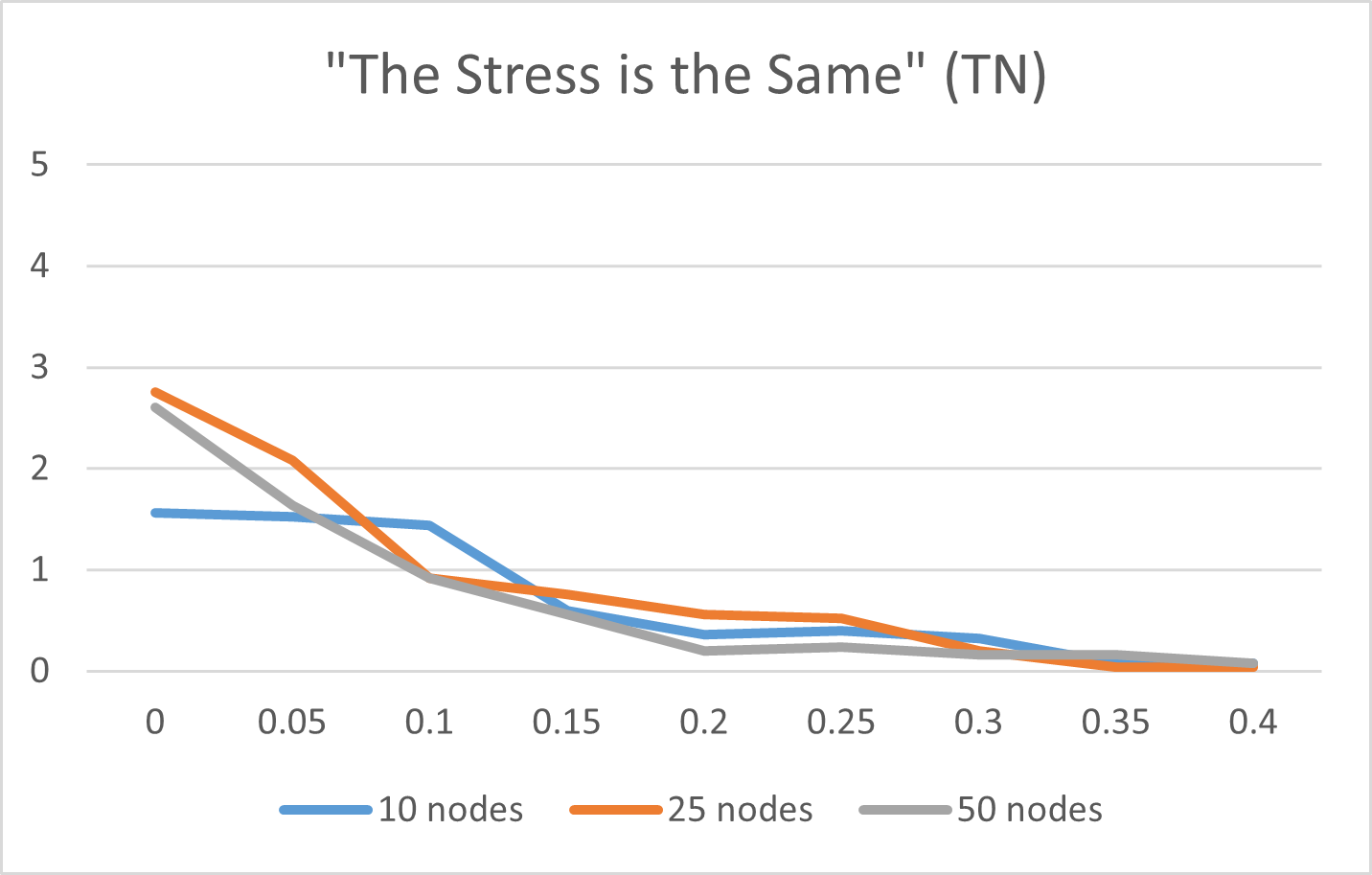}
    \caption{Delta trends for Trained Novices.}
    \label{fig:delta_tn}
\end{figure}

We consider each graph size in turn, plot the average accuracy, response time, and confidence over all 25 participants, and consider trends with respect to delta (Figure~\ref{fig:delta_tn}). There were five trials per delta value, so accuracy is between 0 and 5; high confidence is recorded as 2 for each trial (so confidence per delta is between 0 and 10); time is measured in seconds. There are eight response time data points for which the response time was over 200 seconds (six for n=50, two for n=25). All other 3,367 response time data points were less than 140 seconds. These eight data points (0.24\% of the total data points) were replaced with the mean of the relevant participants' other responses. Participants were given the option to indicate that the stress in the two drawings was the same. Only five of the forty-five trials were pairs with zero delta. Use of this option for other deltas indicates difficulty in distinguishing stress.

\subsubsection{Overall accuracy}
\label{sec:results3}

\begin{figure}[ht]
    \centering
    \begin{subfigure}[b]{0.49\textwidth}
        \centering
        \includegraphics[width=\textwidth]{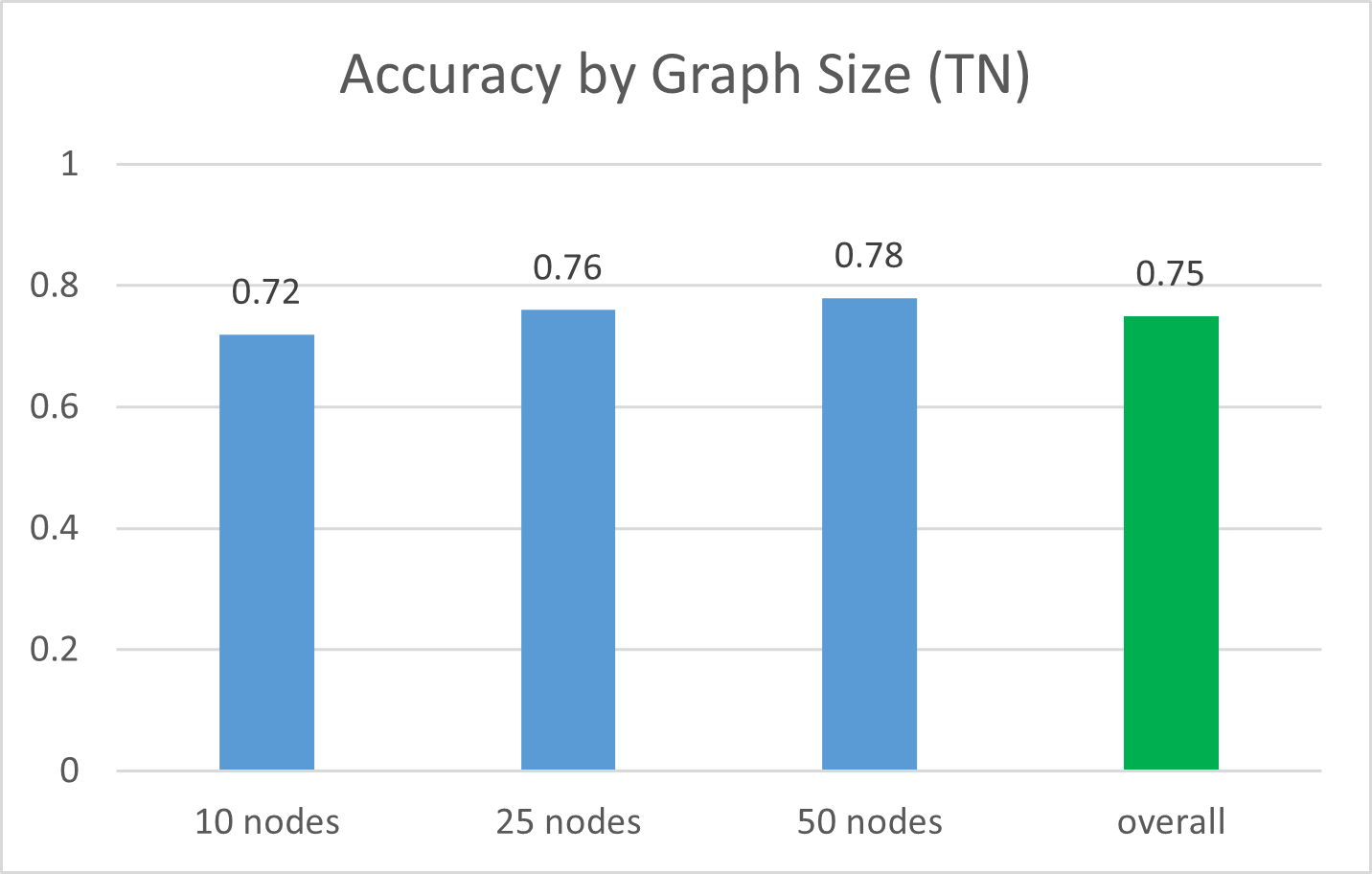}
        \subcaption{\centering}
        \label{fig:overall_tn}
    \end{subfigure}
    \hfill
    \begin{subfigure}[b]{0.49\textwidth}
        \centering
        \includegraphics[width=\textwidth]{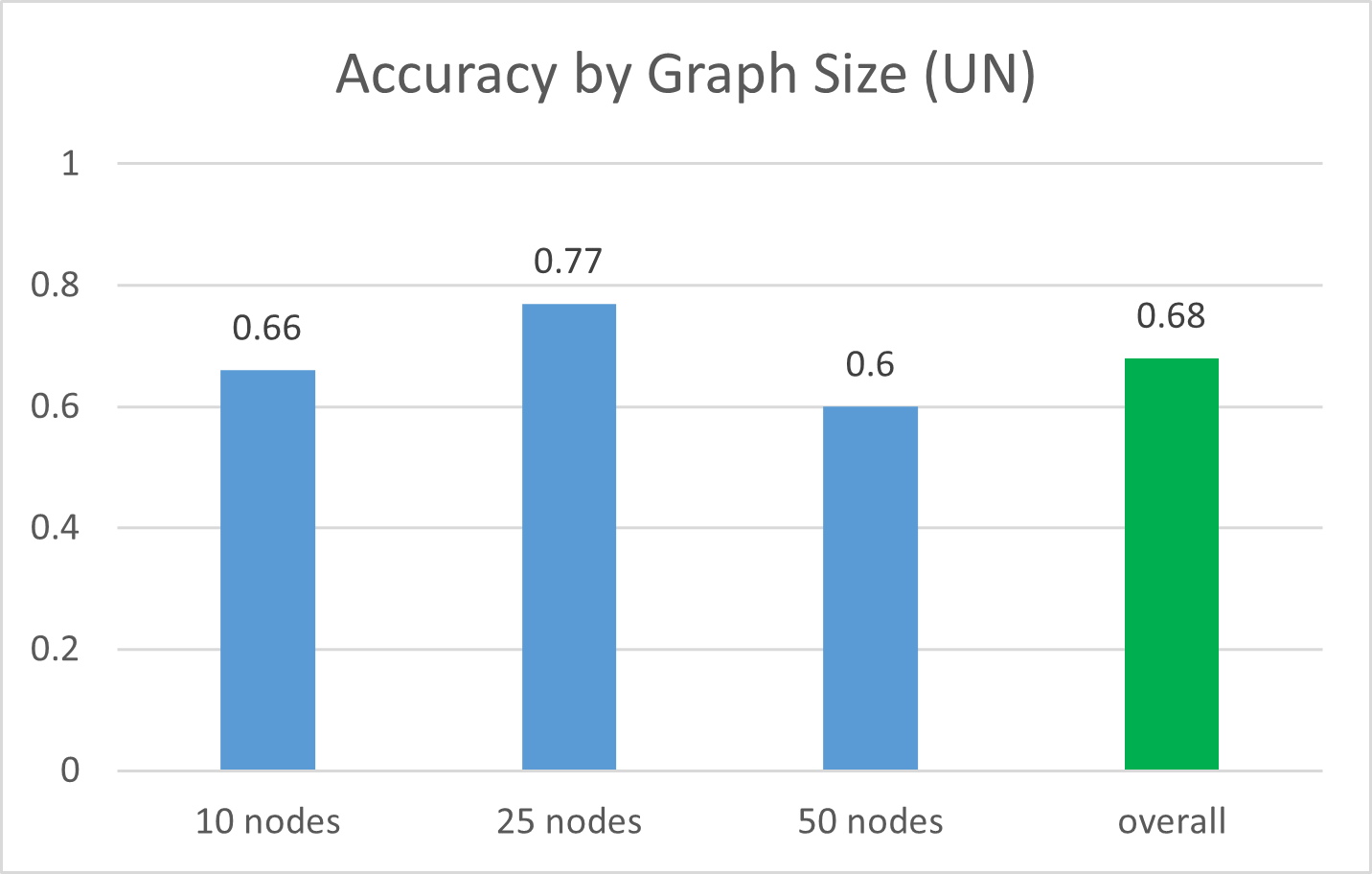}
        \subcaption{\centering}
        \label{fig:overall_un}
    \end{subfigure}
    \hfill
    
    \begin{subfigure}[b]{0.49\textwidth}
        \centering
        \includegraphics[width=\textwidth]{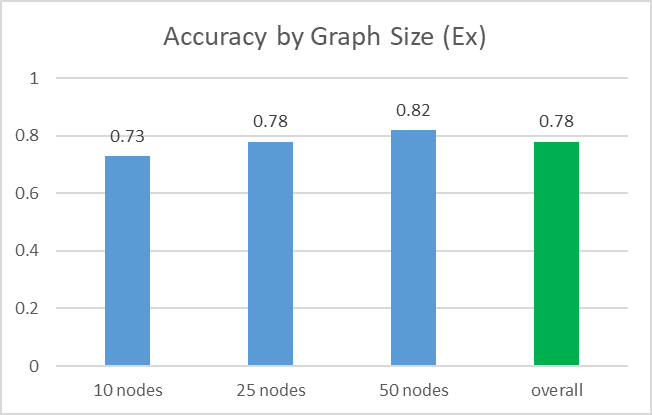}
        \subcaption{\centering}
        \label{fig:overall_ex}
    \end{subfigure}
    \caption{Overall accuracy over all trials for each participant type, with respect to graph size.}
    \label{fig:accuarcy_by_graph_size}
\end{figure}

In this case, we ignore the delta values, and focus on whether the size of the graph makes a difference to the overall mean accuracy of stress perception (Figure~\ref{fig:overall_tn}). An independent measures t-test between the data for 10 nodes and that for 50 nodes reveals a p-value of 0.0495---a barely significant difference.

34 of the 75 participants (45\%) said they found the task difficult; only three said that it was `very easy'.

\subsubsection{Discussion: Performance}

With an overall accuracy rate of 75.2\% (considerably higher than chance: 33\% for three-way multiple choice responses), our data shows that novices who have had the notion of stress explained to them, and have had the opportunity to be trained with feedback, can indeed `see' stress. Contrary to our expectations, the size of the graph does not make the task more difficult, and we see expected trends in accuracy, response time and confidence with respect to difficulty  (as measured by the delta). While there is some uncertainty with the lower deltas of 0, 0.05 and 0.1, by the time the delta reaches 0.15, participants are getting 80\% of the tasks correct, rising as high as 96\% accuracy for 25 nodes with delta=0.4. Looking at the average number of times that participants judged the stress to be the same, we can see that any ambiguity caused by deltas of 0.05 and 0.01 diminishes when the delta increases to 0.15. 
We were surprised by these quantitative results. We had assumed that the complexity of the notion of stress would be difficult for novices to grasp easily, and that it would be more difficult to determine in larger graphs.

\subsubsection{Discussion: Strategies and Perception}

We had asked the participants to explain their strategy in determining the stress differences. Many participants (24) said that they looked at the distances between the nodes, or the distribution of the nodes (5). Some were more specific, noting that they looked at the length of the edges between the nodes (11). Apart from some references to ``messiness/busy-ness'' (10), other features highlighted were symmetry (2), edge crossings (5) or edge closeness (4), angles (5) and ``clusters'' (5). Only one participant made a value judgement on the form of the drawings: ``I choose the one that made me feel less stressed i.e the one that was easiest to understand''. Other notable text summaries include: ``The tightness of the angles, the grouping and how close it was, how ``spiky'' or chaotic the diagrams looked'', ``the ones that looked more erratic and all over the place felt more stressful'' and ``if the lines looked messy, it had more stress.''

Our trained novices were therefore able to perceive `stress' in a graph drawing, using the perception of edge length, node distribution and visual edge density (i.e. the closeness and compactness of edges, including edge crossings) as proxy measures.

\subsection{Untrained Novices (UN)}


Our positive results for trained novices clearly depended on the nature and extent of the training offered to them: not only were they given a detailed written explanation of the stress measures (with examples), they had nine opportunities to attempt the task, with feedback given as to the correct answer for each.   We conducted the experiment again, for 10 participants for each size graph, and omitting the feedback.

Of the 30 participants, there were 16 women, 13 men and 1 unknown. The median age range was 26-35, with 7 participants between 18-25, and four between 56-65. No-one claimed to be `very familiar' with graph drawing; 24 claimed novice status (at most `not very familiar').

\begin{figure}[ht]
    \centering

    \includegraphics[width=0.49\textwidth]{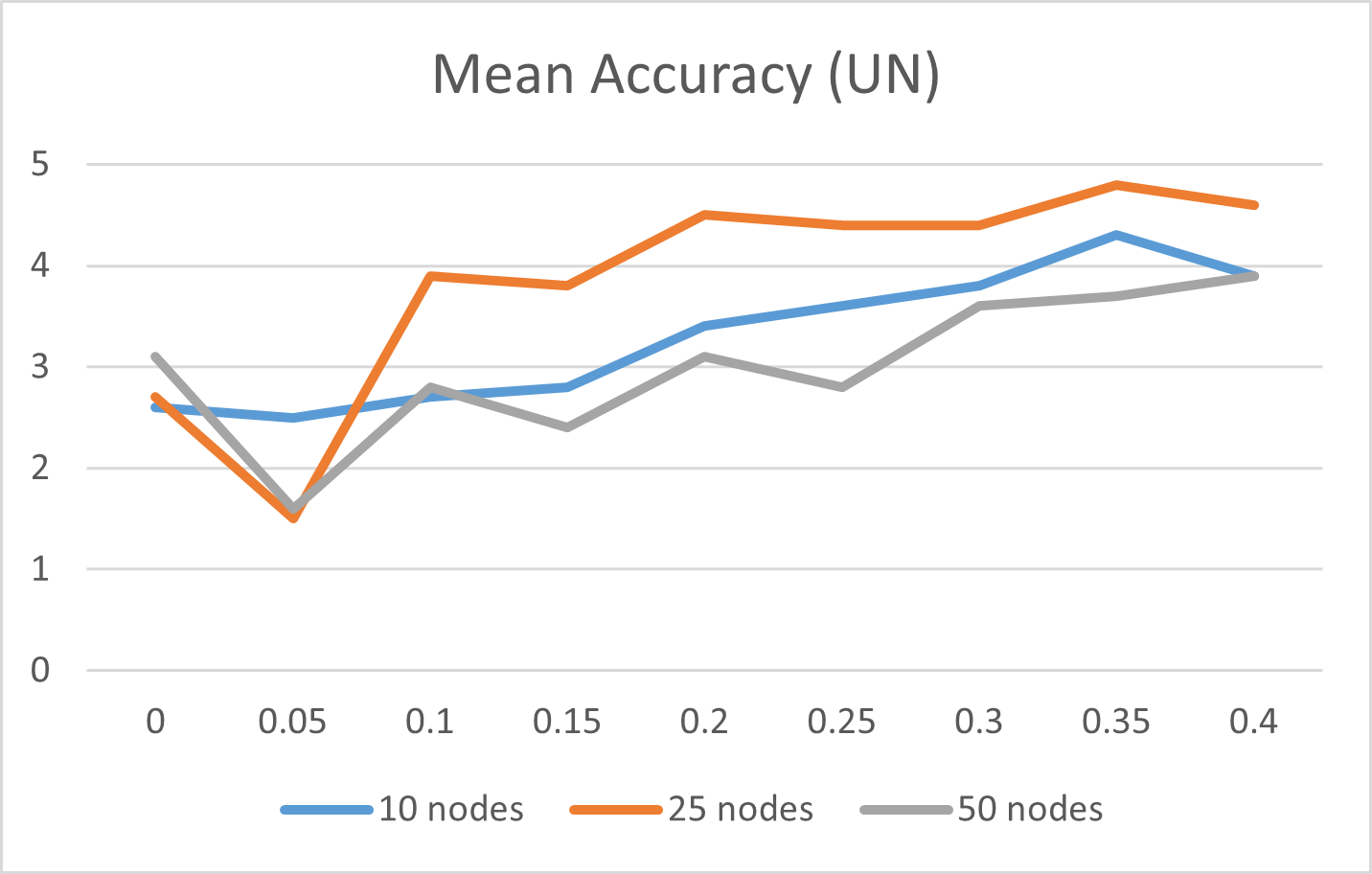}
    \includegraphics[width=0.49\textwidth]{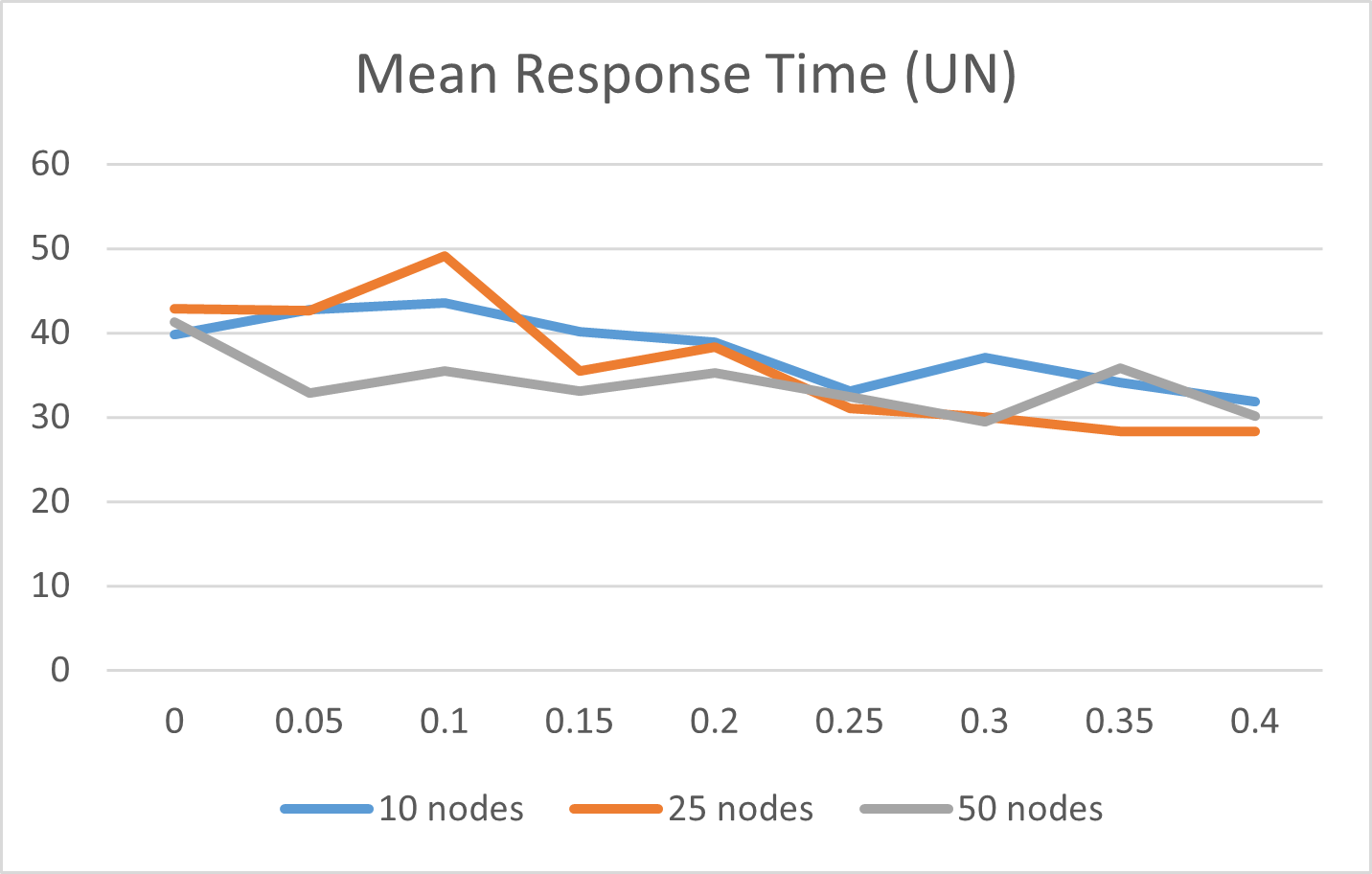}

    \includegraphics[width=0.49\textwidth]{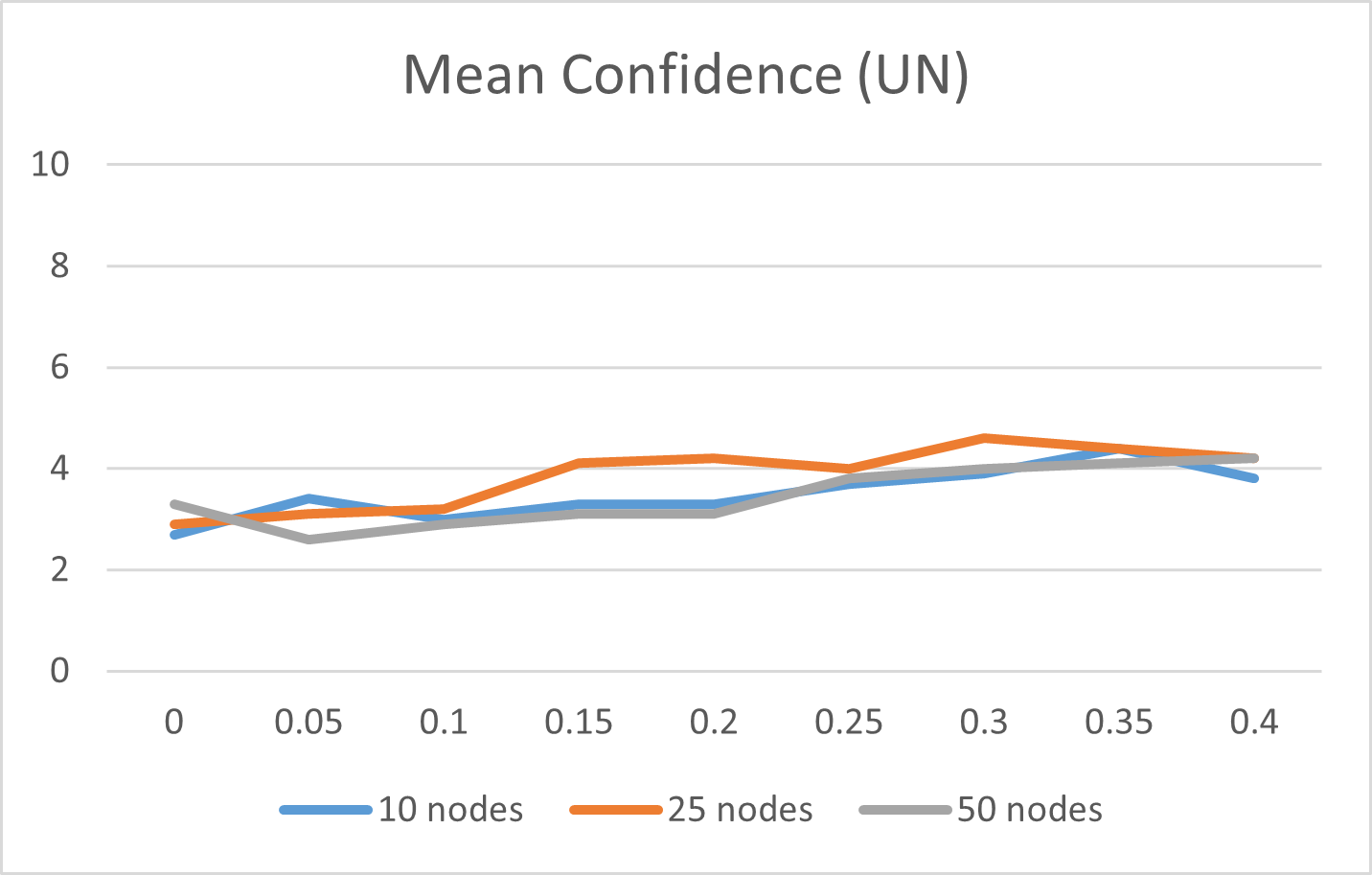}
    \includegraphics[width=0.49\textwidth]{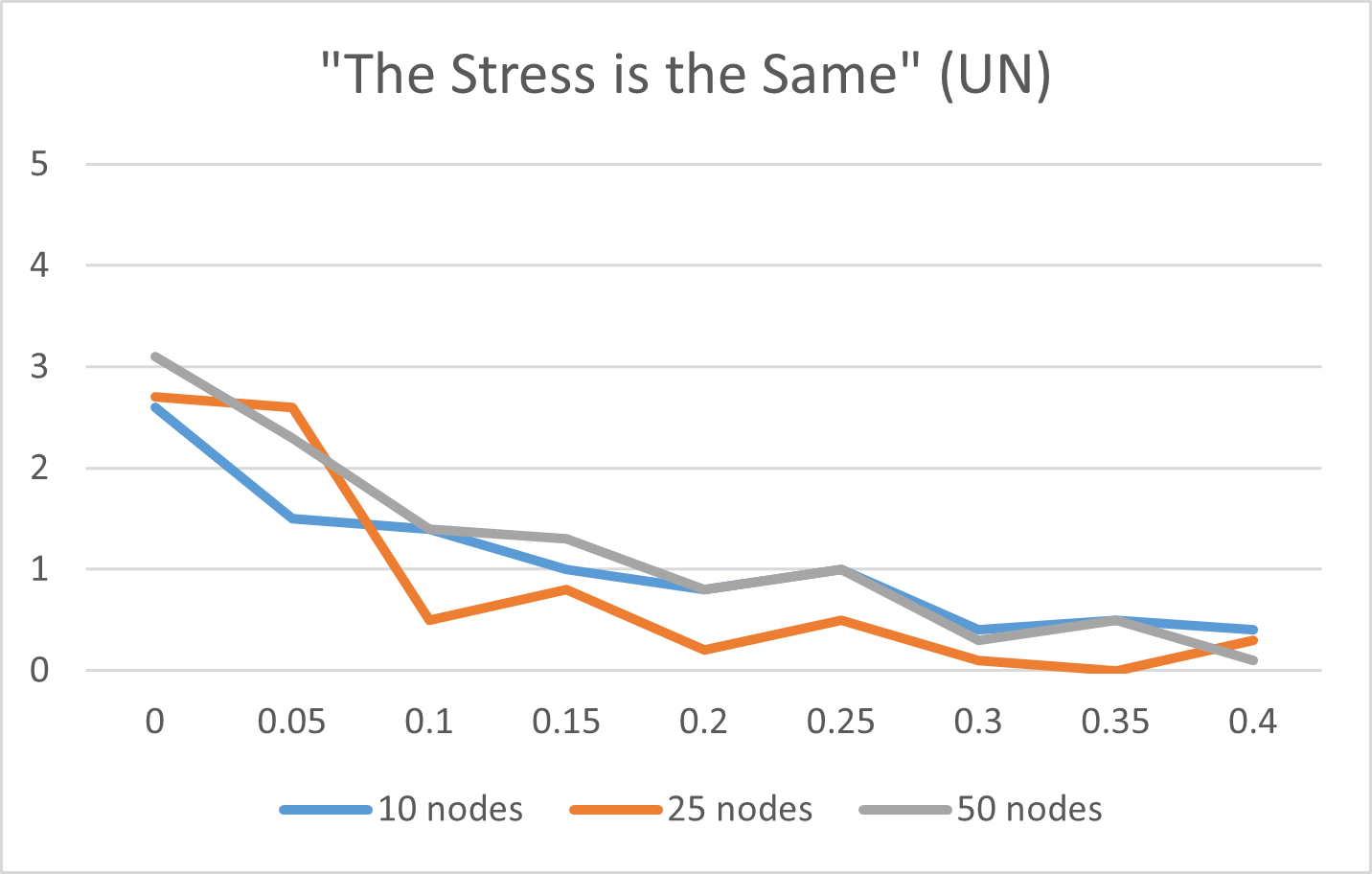}    
    
    \caption{Delta trends for Untrained Novices.}
    \label{fig:delta_un}
\end{figure}

Figure~\ref{fig:delta_un} shows the data for the untrained participants; Figure~\ref{fig:overall_un} shows the overall accuracy. An independent measures t-test between the data for 25 nodes and that for 50 nodes reveals a p-value of 0.0561---an insignificant difference. 23 of the 30 participants said that they found the task difficult.

\subsubsection{Discussion}

The overall trend pattern for the untrained novices is comparable to that of the trained novices: any ambiguity when delta is 0.05 or 0.1 is resolved by the time it increases to 1.5. It appears that the medium-sized graph (n=25) was easier; it is not clear why this might be the case.

The strategies used by these participants were similar to those used by the trained participants: distance between nodes and node distribution (14), `chaotic' (5), and edge lengths (3). There were three references to `open space' (not mentioned by the trained participants): ``Anything that looked congested or compact I found stressful to look at where as when it was more spaced with no sudden lines pointing out it looked much more relaxed.''  Two participants referred to analogies: ``it reminded me of chemical and biological bonds, those with random arrangements where the links/bonds were more likely to break, be fragile, have weaker chain strengths. Those with consistent links/bonds where the lengths and widths are the same were stronger and less likely to separate''; ``some looked like animals, which was less stressful to me. if there was a rogue line jutting out of somewhere I found this stressful. when they had more space I found it less stressful.''

The overall accuracy of the untrained novices was 67.6\% - still greater than chance (33\%) but less than the accuracy for the trained novices (75.2\%). An independent-samples t-test between the overall mean accuracy for trained and untrained novices gives a p-value of 0.009; thus, the additional training made a significant difference to the performance of the novice.

\subsection{Experts}
Having gathered data on how well novices can perceive stress after being presented with a rudimentary explanation of the concept, we were interested in how graph drawing experts (who already know the concept well) would fare in the same experiment. Eight graph drawing experts---known colleagues of the authors (7M, 1F, median age 36-45)---completed all the trials for all three graph sizes. The experts were shown the same written explanation of `stress', but were not given any training.

\begin{figure}[ht]
    \centering

    \includegraphics[width=0.49\textwidth]{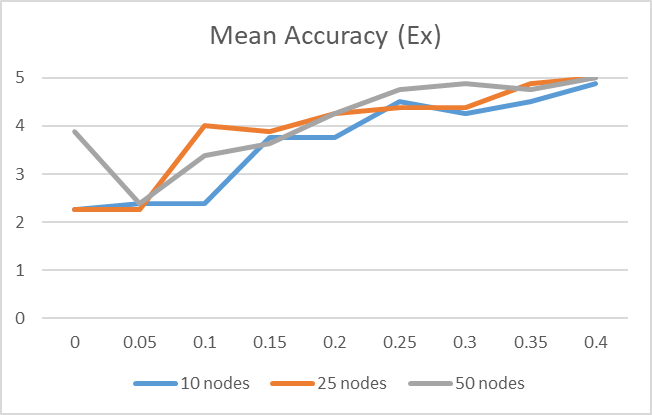}
    \includegraphics[width=0.49\textwidth]{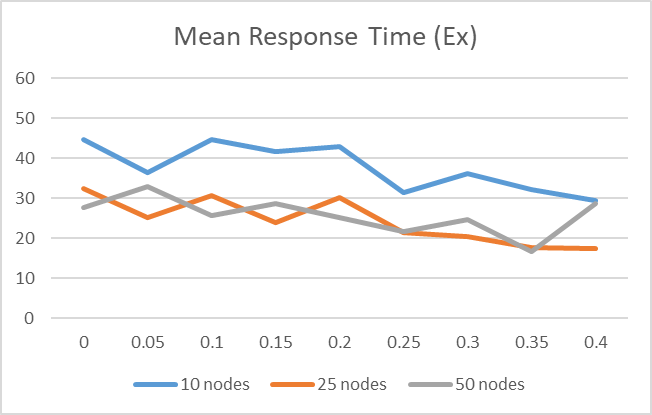}

    \includegraphics[width=0.49\textwidth]{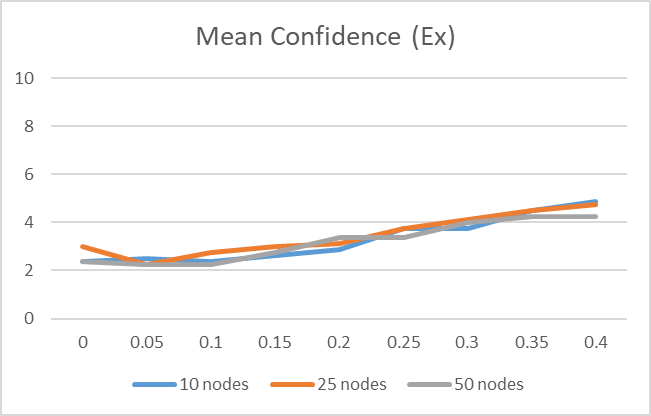}
    \includegraphics[width=0.49\textwidth]{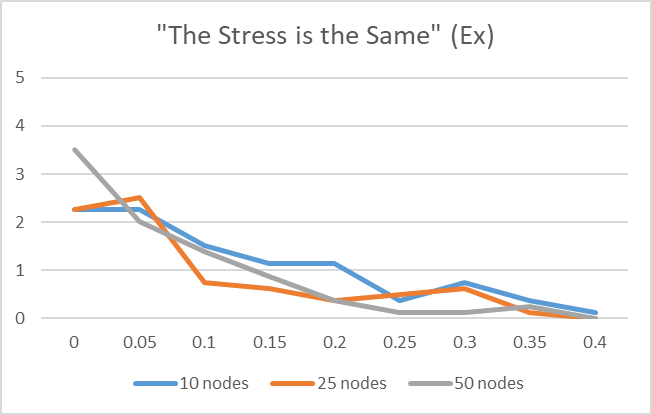}
    
    
    \caption{Delta trends for Expert participants.}
    \label{fig:delta_ex}
\end{figure}

Figure~\ref{fig:delta_ex} shows the data for the expert participants; Figure~\ref{fig:overall_ex} shows the overall accuracy. A repeated measures t-test between the data for 10 nodes and that for 50 nodes reveals a p-value of 0.089---an insignificant difference.

\subsubsection{Discussion}

The trend patterns for the experts are remarkably similar to the novices. As with both trained and untrained novices, confusions between the pairs with deltas of 0.1 and less are resolved for those with values of 0.15 and above. The longer response time for the graphs with 10 nodes can be explained by the fact that these were the trials that all experts completed first, before they then looked at the larger graphs (suggesting a `warming-up' period). The experts' confidence is remarkably low, closely matching that of the novices.

\begin{figure}[h]
    \centering
    \includegraphics[width=0.45\textwidth]{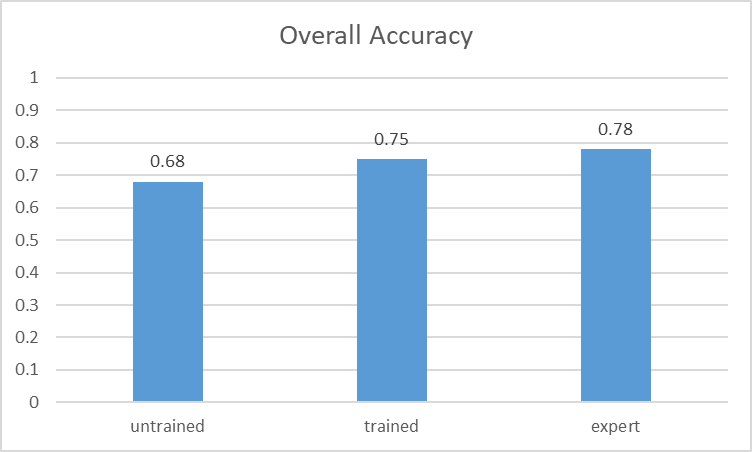}
    \caption{Overall accuracy over all trials for each participant group.}
    \label{fig:overall_participant_acuracy}
\end{figure}

The strategies used by the experts included line length (2), distances between nodes (2), and edge crossings (2). There were, however, some less specific considerations and some clear value judgements: ``drawings with Low stress just look ``right''... high stress drawings look more random showing features that a human graph drawer would attempt to correct in a second iteration''; ``general feeling how much the nodes are settled into place''; ``how well the overall structure is highlighted''; ``how well untangled they are.'' It appears that experts do not (like novice) need to rely so much on visual proxies, but their experience gives them an intuitive `feel' for stress.
This `feel' is, however, not fail-safe, with the experts having only an overall accuracy of 77.5\%. Figure~\ref{fig:overall_participant_acuracy} shows the relative accuracy of the three categories of participant. 

An independent-samples t-test between the overall mean accuracy for untrained and trained novices gives a p-value of 0.009; thus, the additional training made a significant difference to the performance of the novices.

An independent-samples t-test between the overall mean accuracy for trained and expert gives a p-value of 0.361; thus, there is no significant difference in accuracy between trained novices and experts. There is also no significant difference between mean response time for trained novices (7.89s) and experts (5.86s); p=0.065.

We speculate that trained novices thought more carefully about their choices, using visual proxies they had identified during the training sessions, while the experts made rapid `gut-feel' decisions. Despite this, all participants reported low confidence in their responses---even the experts.

\subsection{Correlations Between Stress and Other Metrics}
\label{sec:corrs}
\begin{figure}[ht]
    \centering
    \includegraphics[width=\textwidth]{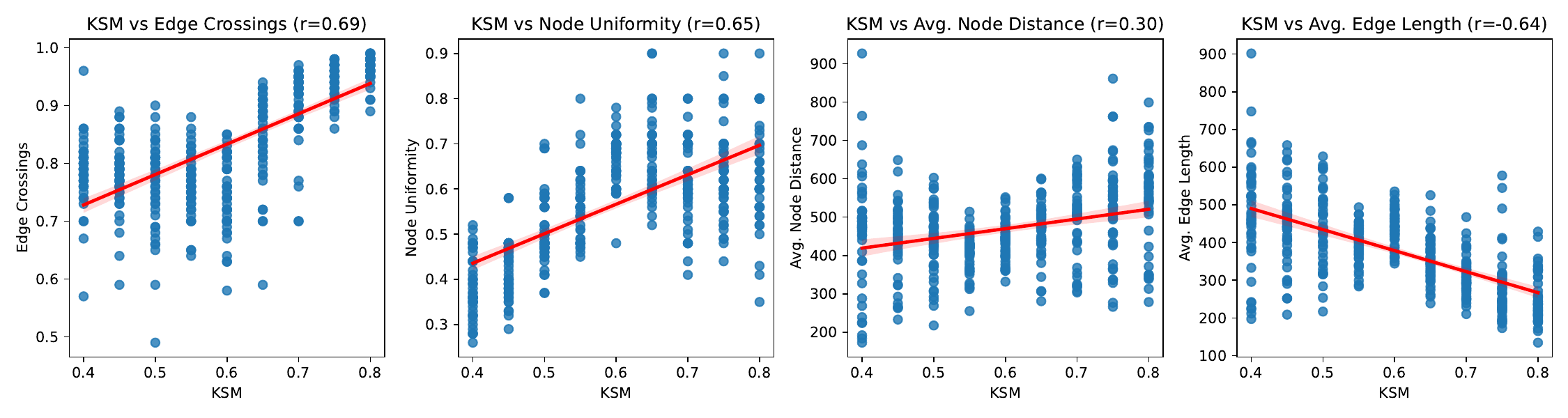}
    \caption{Correlations between KSM and other graph drawing metrics.}
    \label{fig:corrs}
\end{figure}

We calculated the correlations between KSM and four other visual properties present in the `strategy' responses: edge crossings, distribution of nodes in the drawing space, average distance between nodes, and average edge length (Figure~\ref{fig:corrs}). For the distribution of nodes and edge crossings, we use the Node Uniformity and Edge Crossing metrics defined by Mooney et al.~\cite{mooney2024multi}. We compute these metrics on our set of 405 stimuli drawings.

We find that KSM is positively correlated with the average distance between nodes (0.30), the distribution of nodes (0.65), and the number of edge crossings (0.69). KSM is negatively correlated with average edge length (-0.64). Recalling that KSM values closer to 1.0 correspond to lower stress, we can make the following observations:
\begin{itemize}
    \item Drawings with nodes which are farther apart on average have lower stress. 
    \item Drawings which have shorter edge lengths on average have lower stress.
    \item Drawings in which the nodes are more evenly distributed around the drawing space have lower stress.
    \item Drawings with fewer edge crossings have lower stress.
\end{itemize}

These observations suggest that some of the participants' `strategies' are effective visual proxies for stress, though do not fully capture it.


\section{Conclusion}
\label{sec:conculsion}

These experiments represent the first investigation of whether people can `see' stress, an invisible property of graph drawings that is defined by both the geometric form of the graph drawing as well the underlying structure of the graph. We find that it is indeed possible to describe the notion of stress to people unfamiliar with graphs and graph drawing in manner that allows them to perceive it. Unprompted, they are able to devise their own visible proxies for the invisible stress feature---that is, they easily identify appropriate visual features of drawings which can be used to assess stress: in particular the geometric length of edges, distances between nodes and node distribution, and `compactness'/ `clustering'/ `density'. We can't be sure that the participants utilised the exact geometric definition of stress but their performance suggests that they were able to perceive its visual features. A few participants referred to the personal stress that they felt while looking at the drawings, thus transferring a mathematical concept into an emotion.

Surprisingly, experts do not perform much better than trained novices. While there is some evidence of them also using visual proxies, they also appear to rely on high-level concepts like `untangling', `general feeling' and `settled nodes'---more than is evident in the strategies described by the novices.

Our findings do not differ with respect to graph size. We were surprised at this, but this could be explained by the fact that the participants focused on these overall visual proxies rather than the edges themselves. Our results therefore appear to be generalisable to graph drawings of different sizes.

This work is important for any human experimental work on the perception and use of graph drawings where stress is deemed an important feature. Not only have we shown that people can understand the concept of stress sufficiently well to be able to perceive it, we have also identified those visual features that they consider to be most related to stress: edge length, node distribution and concentrated areas of edges and edge crossings. Future experimental work can now build on these results to investigate, for example, nuances of different implementations of stress, trade-offs between stress and other layout principles, the smallest delta for stress perception, and the form of user-generated drawings where low stress is an explicitly stated goal.

\subsection{Future Work}
\label{sec:future_work}
All empirical studies are limited by necessary practical parameters. In this case, the stimuli are small and sparse graphs. A follow up experiment on larger and denser graphs would inform on how the perception of stress varies as graph sizes increase. While our results show only a small difference in terms of perception between graphs of size 10 and 50, this may not be the case in general as the size of the graph increases. However, as the size and density of the graph increases, producing drawings with low stress becomes increasingly challenging. 


Our results suggest that stress deltas of 0.1 and below are confusing, but that stress differences of 0.15 and above are more discernible. These observations are also limited by the range of KSM values used in the experiment. It may be the case that the discernible difference in KSM is non-uniform; i.e., the difference between drawings with KSM of 0 and 0.15 may be more or less discernible than that of 0.85 and 1.0. A more comprehensive `Just Noticeable Difference' methodology (such as that used by Soni et al. to measure the perception of graph density~\cite{soni_jnd}) could validate this, and perhaps result in a more specific threshold. This methodology would allow us to see what the smallest perceptible difference in stress is, and see if this follows Weber's law. This requires careful creation of a broader set of stimuli as well as a larger number of participants.

Investigation into additional `non-visual' metrics, which are not associated with the principles of visual perception, such as Neighbourhood Preservation, would provide insights into their perceptibility. This may also inform on the usefulness of such metrics as criteria for optimisation and/or evaluation of graph drawings.




\bibliography{stress}
\clearpage
\appendix

\section{Stress Explanation}
\label{sec:appendix_stress}
\noindent
\textbf{Definitions: networks and network drawings}\\
A network is made up of objects and connections. For example, this social network depicts people (represented as circles) and friendships (represented as lines between the circles). Amy has four friends; Ted has two.

\begin{figure}[H]
    \includegraphics[width=0.2\textwidth]{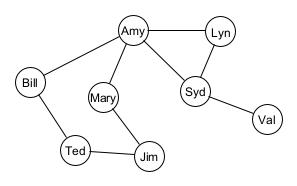}
\end{figure}
\noindent
The same network can be drawn in many different ways by changing the position of the objects. For example, here are four drawings of the same network.

\begin{figure}[H]
    \centering
    \begin{subfigure}[b]{0.2\textwidth}
        \centering
        \includegraphics[width=\textwidth]{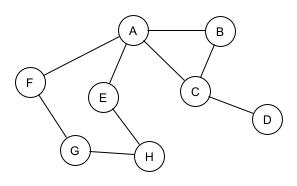} 

    \end{subfigure}
    \hfill
    \begin{subfigure}[b]{0.2\textwidth}
        \centering
        \includegraphics[width=\textwidth]{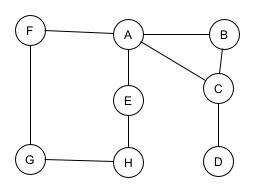}

    \end{subfigure}
    \hfill
    \begin{subfigure}[b]{0.2\textwidth}
        \centering
        \includegraphics[width=\textwidth]{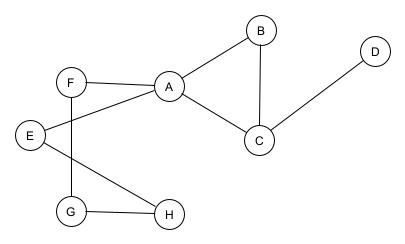}

    \end{subfigure}
    \hfill
    \begin{subfigure}[b]{0.1\textwidth}
        \centering
        \includegraphics[width=\textwidth]{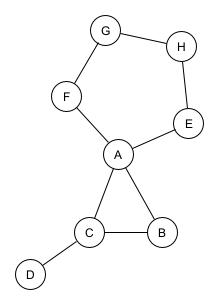} 
    \end{subfigure}
\end{figure}

\noindent
A `path' is a series of steps between objects. For example, in the network below, the length of the path between G and F is 4; the length of the path between A and D is 2 or 3 (depending on whether you go through B or not).

\begin{figure}[H]
    \includegraphics[width=0.2\textwidth]{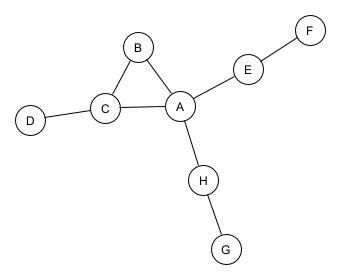}
\end{figure}

\noindent
The `shortest path' is, as its name suggests, the shortest path when there is more than one way to get from one object to another.
\\ \\
For example, in the network below, the shortest path between F and B is 2 (going through A, but not C or G/H/E); the shortest path between D and H is 4 (going through C/A, but not through B or F/G).

\begin{figure}[H]
    \includegraphics[width=0.2\textwidth]{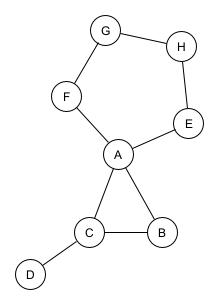}
\end{figure}

\noindent
\textbf{Definitions: visual properties of network drawings}\\
Given that there is more than one way to draw a network, we can distinguish between them by their ‘visual properties’.
\\ \\
For example, the drawing on the left has ‘tighter angles’ than the one on the right. These are both drawings of the same network.

\begin{figure}[H]
    \centering
    \begin{subfigure}[b]{0.2\textwidth}
        \centering
        \includegraphics[width=\textwidth]{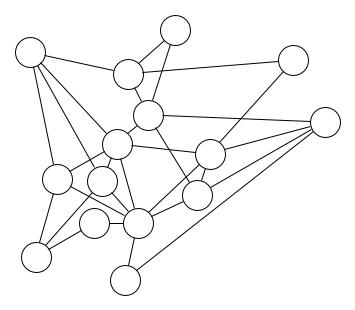} 

    \end{subfigure}
    \hfill
    \begin{subfigure}[b]{0.2\textwidth}
        \centering
        \includegraphics[width=\textwidth]{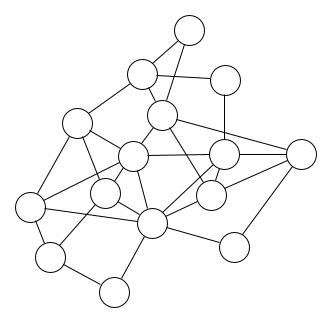}

    \end{subfigure}

\end{figure}

\noindent
And the drawing on the right has ‘more symmetry’ than the one on the left. These are both drawings of the same network.

\begin{figure}[H]
    \centering
    \begin{subfigure}[b]{0.2\textwidth}
        \centering
        \includegraphics[width=\textwidth]{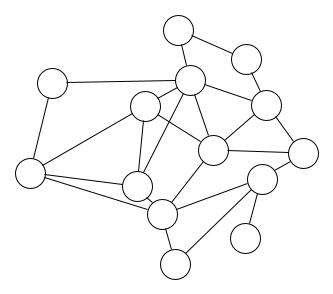} 

    \end{subfigure}
    \hfill
    \begin{subfigure}[b]{0.2\textwidth}
        \centering
        \includegraphics[width=\textwidth]{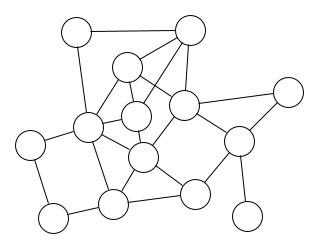}

    \end{subfigure}

\end{figure}

\noindent
In this experiment, we are interested in the visual property of `stress'.
\\ \\
A drawing has \textbf{low stress} if the distance between pairs of objects is proportional to the length of the shortest path between them.
\\ \\
In its simplest form, the following network drawing has very low stress: the distance between each pair of objects is directly proportional to the (shortest) path between them.

\begin{figure}[H]
    \includegraphics[width=0.3\textwidth]{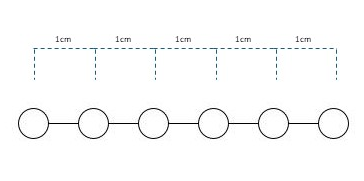}
\end{figure}

\noindent
We just need to move one of the objects to \textbf{increase the stress}---the distance between the two objects at each end (6cm) is now longer proportional to the length of the path between them (5).

\begin{figure}[H]
    \includegraphics[width=0.25\textwidth]{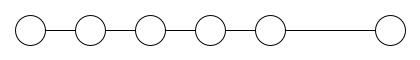}
\end{figure}

\noindent
The same simple network can be drawn with even \textbf{higher stress}, where there is barely any relationship between the distance between the objects and the length of the paths between them.

\begin{figure}[H]
    \includegraphics[width=0.1\textwidth]{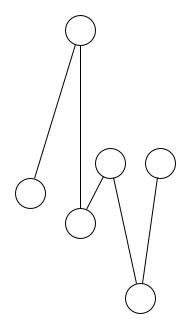}
\end{figure}

\noindent
Similarly, here is another network with very low stress, with two versions of higher stress.

\begin{figure}[H]
    \centering
    \begin{subfigure}[b]{0.15\textwidth}
        \centering
        \includegraphics[width=\textwidth]{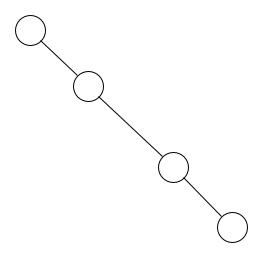} 

    \end{subfigure}
    \hfill
    \begin{subfigure}[b]{0.15\textwidth}
        \centering
        \includegraphics[width=\textwidth]{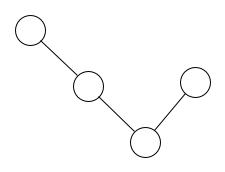}

    \end{subfigure}
    \hfill
    \begin{subfigure}[b]{0.15\textwidth}
        \centering
        \includegraphics[width=0.15\textwidth]{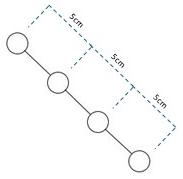}
    \end{subfigure}

\end{figure}

\noindent
Of course, this more-or-less-stress judgement becomes more difficult with larger networks. Here are some more examples showing the same networks drawn with different amounts of stress. In all cases, the network on the left has less stress than the network on the right. Thus, the network on the left maintains the distance/path relationship between pairs of objects better than the one on the right.

\begin{table}[H]
\centering
\begin{tabular}{|l|l|} 
\hline
\multicolumn{1}{|c|}{lower stress} & \multicolumn{1}{c|}{higher stress}  \\ 
\hline
\includegraphics[width=0.1\textwidth]{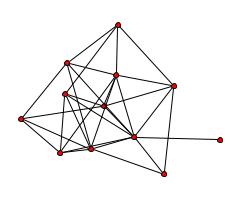}                                   &  \includegraphics[width=0.2\textwidth]{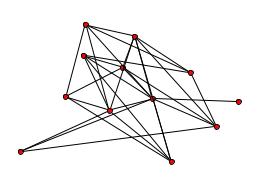}                                     \\ 
\hline
\includegraphics[width=0.1\textwidth]{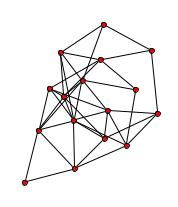}                                     &  \includegraphics[width=0.2\textwidth]{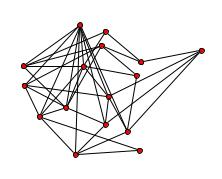}                                     \\ 
\hline
 \includegraphics[width=0.1\textwidth]{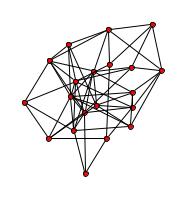}                                    & \includegraphics[width=0.2\textwidth]{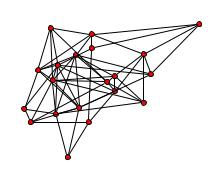}                                      \\ 
\hline
\includegraphics[width=0.1\textwidth]{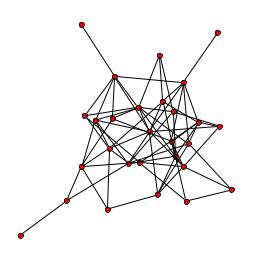}                                     & \includegraphics[width=0.2\textwidth]{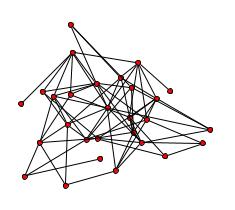}                                      \\
\hline
\end{tabular}
\end{table}

\noindent
Of course, we cannot assess the differences in stress by doing all the distance and path length calculations in our head! But we can get a ‘feeling’ as to when one drawing has less stress than another.
\\ \\
Before you start the experiment, we will ask you to make your own judgements, and let you know whether you are correct or not.

\section{Metric Formulae}
This section describes the metrics used in Section~\ref{sec:corrs}. Mooney et al.~\cite{mooney2024multi} provide definitions of ten graph drawing metrics: we use the Edge Crossings and Node Uniformity metrics. All metrics were calculated using Python and the NetworkX library. The average distance between nodes and average edge length can be trivially computed by iterating over the pairs of nodes.

\subsection{Node Uniformity}
Node Uniformity measures how uniformly distributed nodes are inside the bounding box of the drawing. We calculate it by splitting the box into cells and counting the number of nodes in each cell, then comparing this to an ideal distribution. The size of cells (and hence number of cells) is scaled by the square root of the number of nodes in the graph.

\subsection{Edge Crossings}
The number of edge crossings in the drawing, scaled against the total possible crossings. 

    \begin{equation}
        EC = 1 - \begin{cases}
                    \frac{c}{c_{mx}}, & \text{if $c_{mx}>0$}\\
                    0, & \text{otherwise}
                  \end{cases}
    \end{equation}

    Where $c$ is the number of crossings and $c_{mx}$ is the upper bound on the number of possible crossings,

    \begin{equation}
        c_{mx} = \frac{m(m-1)}{2} - c_{deg}
    \end{equation}

    \begin{equation}
        c_{deg} = \frac{1}{2} \sum_{j=1}^{n} degree(u_j)(degree(u_j)-1)
    \end{equation}

    and $c_{deg}$ is the number of crossings which are impossible due to the fact that adjacent edges cannot cross.

    We expand upon this definition to reduce $c_{mx}$ further by accounting for triangles and 4-cycles in the graph. Pairs of triangles can only cross at most six times, as opposed to the nine calculated by only using $c_{deg}$. We do however have to account for triangles with shared edges and nodes, as these cases are partially handled by $c_{deg}$. Additionally non-adjacent edges to a triangle can only cross at most two of the triangle's edges. We call the number of crossings which are impossible due to triangle interactions $c_{tri}$ which is calculated using \cref{alg:triangles}.

    \begin{algorithm}
    \caption{Algorithm to calculate $C_{tri}$}\label{alg:triangles}
    \begin{algorithmic}
    \Procedure{Reduce Triangles}{$Graph: G$}
    \State $c_{tri} \gets 0$
    \For{each triangle, t, in G}
        \For{each edge, m, in G}
            \If{m is not part of, or adjacent to any triangle in G}
                \State $c_{tri} \gets c_{tri} + 1$
            \EndIf
        \EndFor
    \EndFor
      
    \For{each pair of triangles, t, u, in G}

        \If{t and u share an edge}
            \State $c_{tri} \gets c_{tri} + 1$
        \ElsIf{t and u share a node}
            \State $c_{tri} \gets c_{tri} + 2$
        \Else{}
            \State $c_{tri} \gets c_{tri} + 3$
        \EndIf

    \EndFor
    \EndProcedure
    \end{algorithmic}
    \end{algorithm}

    We can also reduce $c_{mx}$ by the number of 4-cycles in the graph, $c_{4cyc}$, due to the fact that if two edges in a 4-cycle cross, it is impossible for the other two edges to cross.

    The final calculation for $c_{mx}$ becomes:

    \begin{equation}
        c_{mx} = \frac{m(m-1)}{2} - c_{deg} - c_{tri} - c_{4cyc}
    \end{equation}

    Giving a tighter upper bound on the number of possible crossings.

\end{document}